\def\rfr#1{Equation~(\ref{#1})}
\def\rfrs#1#2{Eqs.~(\ref{#1})--(\ref{#2})}

\def\dert#1#2{\frac{{{\mathrm{d}}}{#1}}{{{\mathrm{d}}}{#2}}}

\def\virg#1{``#1"}

\def\eqi{\begin{equation}}
\def\eqf{\end{equation}}
\def\eqia{\begin{eqnarray}}
\def\eqfa{\end{eqnarray}}

\def\rp#1#2{{#1\over#2}}
\def\lb#1{\label{#1}}

\def\bds#1{\boldsymbol{#1}}

\def\ton#1{\left(#1\right)}
\def\qua#1{\left[#1\right]}
\def\grf#1{\left\{#1\right\}}

\documentclass[onecolumn]{aastex}
\usepackage{morefloats}
\usepackage[title]{appendix}
\usepackage{hyperref}
\usepackage{booktabs}
\usepackage[table,xcdraw]{xcolor}
\usepackage{multirow}
\usepackage{rotating,tabularx}
\usepackage{float}
\usepackage{enumerate}
\usepackage{rotating}
\usepackage[polutonikogreek,english]{babel}
\usepackage{amsmath,starfont,textgreek,w-greek,wasysym}
\usepackage{amsthm}
\usepackage{amscd,lineno}
\usepackage{amssymb,dsfont}
\usepackage{graphicx,epsfig}
\usepackage{txfonts}
\usepackage{xr-hyper}

\RequirePackage{color}

\newcommand{\emaila}{lorenzo.iorio@libero.it}

\allowdisplaybreaks[1]

\begin{document}

\title{Effects of General Relativistic Spin Precessions on the Habitability of Rogue Planets Orbiting Supermassive Black Holes}
\shortauthors{L. Iorio}
\author{Lorenzo Iorio\altaffilmark{1} }
\affil{Ministero dell'Istruzione, dell'Universit\`{a} e della Ricerca
(M.I.U.R.)-Istruzione
\\ Permanent address for correspondence: Viale Unit\`{a} di Italia 68, 70125, Bari (BA),
Italy}
\email{\emaila}

\begin{abstract}
Recently, the possibility that several starless telluric planets may form around supermassive black holes (SMBHs) and receive an energy input from the hole's accretion disk, which, under certain plausible circumstances, may make them habitable in a terrestrial sense, has gained increasing attention. In particular, an observer on a planet orbiting at distance $r=100$ Schwarzschild radii from a maximally rotating Kerr SMBH with mass $M_\bullet = 1\times 10^8\,M_\odot$ in a plane slightly outside the equator of the latter, would see the gravitationally lensed accretion disk the same size as the Sun as seen from the Earth. Moreover, the accretion rate might be imagined to be set in such a way that the apparent disk's temperature would be identical to that of the solar surface. We demonstrate that the post-Newtonian (pN) de Sitter and Lense--Thirring precessions of the spin axis of such a world would rapidly change, among other things, its tilt, $\varepsilon$, to its orbital plane by tens to hundreds of degrees over a time span of, say, just $\Delta t =400\,\mathrm{yr}$, strongly depending on the obliquity $\eta_\bullet$ of the SMBH's spin to the orbital plane. Thus, such relativistic features would have per se a relevant impact on the long-term habitability of the considered planet. Other scenarios are examined as well.
\end{abstract}

Unified Astronomy Thesaurus concepts:{
Black holes (162); Exoplanets (498); Celestial mechanics (211);
Astrobiology (74); Gravitation (661); General Relativity (641)
}

\section{Introduction}
Supermassive black holes (SMBHs) \citep{Melia09} are most likely lurking in the center of nearly all large galaxies. The largest ones for which empirical evidences, collected with different techniques exist are  HOLM 15A, whose mass is $M_\bullet=4.0\times 10^{10}\,M_\odot$ \citep{2019arXiv190710608M}, and TON 618 $\ton{M_\bullet=6.6\times 10^{10}\,M_\odot}$ \citep{2004ApJ...614..547S}. IC 1101, whose mass has not been measured directly, may even reach the $M_\bullet \leq 10^{11}\,M_\odot$ level \citep{2017MNRAS.471.2321D}. A SMBH with $M_\bullet =6.5\times 10^9\,M_\odot$ \citep{2019ApJ...875L...6E}, whose shadow was recently imaged by the Event Horizon Telescope collaboration \citep{2019ApJ...875L...1E}, is located at the center of the supergiant elliptical galaxy M87. Milky Way is believed to host a SMBH with $M_\bullet =4.15\times 10^6\,M_\odot$ \citep{refId0} in Sgr A$^\ast$ at the Galactic Center.

Currently, it is widely accepted that SMBHs  are at the basis of the extremely powerful electromagnetic emission of the active galactic nuclei (AGNs) \citep{1999PNAS...96.4749F} due to the release of gravitational energy of the infalling matter \citep{AGNs}.

It was recently pointed out that the accretion disks around SMBHs in low-luminosity AGNs \citep{2017Natur.549..488R}  may sustain the formation of several telluric, Earth-like rogue (i.e. starless) planets at some parsecs (pc) from them \citep{2019arXiv190906748W} which, under certain circumstances, may even sustain habitable environments \citep{Lingam_2019}. In particular, \citet{2019arXiv190906748W} considered a dusty disk ranging from $r\simeq 0.1\,\mathrm{pc}$ to $r\simeq 100\,\mathrm{pc}$ around SMBHs with masses $M_\bullet\simeq 10^6-10^9\,M_\odot$. \citet{Lingam_2019} considered SMBHs in the mass range $M_\bullet\simeq 10^9-10^{10}\,M_\odot$, and found that planets at subpc distances from the hosting AGNs may become uninhabitable because of complex interactions with the dusty torus, as well as strong outflows and winds from the accretion disk. On the other hand, \citet{Lingam_2019} warned that the $\simeq 1\,\mathrm{pc}$ threshold should not be regarded as a rigid cutoff because of the remarkable heterogeneity among active galaxies. Moreover, our understanding
of their central regions is currently far from definitive. The zones favorable for prebiotic chemistry and photosynthesis extend up to $\simeq 44\,\mathrm{pc}$ and $\simeq 340\,\mathrm{pc}$, respectively \citep{Lingam_2019}. If $M_\bullet\simeq 10^4-10^5\,M_\odot$, such boundaries reduce below the pc scale \citep{Lingam_2019}. Previous, more pessimistic evaluations of the impact of the AGN phase of Sgr A$^\ast$ in our Galaxy can be found in \citet{2017NatSR...716626B}. \citet{2019arXiv191000940S} explored some potentially harmful consequences for life on a rogue planet orbiting a $10^8\,M_\odot$ SMBH just outside its event horizon.\footnote{
\citet{2019arXiv191000940S} took his inspiration from the fictional Miller's planet closely orbiting a SMBH with that mass, named Gargantua, in the movie \textit{Interstellar}.} Interestingly, \citet{2019arXiv191000940S} also noted  that for a planet orbiting a Gargantua-like, maximally spinning SMBH at  $rsimeq 100$ Schwarschild radii\footnote{$G$ is the Newtonian gravitational constant, and $c$ is the speed of light in vacuum. The gravitational radius $r_g$ used by \citet{2019arXiv191000940S} as a unit of length is half the Schwarzschild radius.} $R_\mathrm{s}=2GM_\bullet/c^2$ in a tilted plane, even if only slightly above or below the SMBH's equatorial plane, the gravitational bending of the electromagnetic waves coming from the assumed Novikov--Thorne type accretion disk \citep{NT1973} would make the latter visible as a lensed ring having roughly the same size of the Sun as viewed from the Earth. Furthermore, for a given mass accretion rate, which, from Figure 7 of \citet{2019arXiv191000940S} seems to be around\footnote{Here, $\dot M_\mathrm{Edd}$ is the Eddington accretion rate \citep{2013rehy.book.....R}.} $\dot M\simeq 10^{-9}\,\dot M_\mathrm{Edd}$, it would even be possible to match the apparent blackbody temperature with the Sun's temperature $T\simeq 6000\,\mathrm{K}$ \citep{2019arXiv191000940S}. Another SMBH--planet scenario potentially viable to sustain life was recently investigated by \citet{Bakala_2020}. They found that, for a narrow range of radii of circular orbits very close to an almost maximally spinning SMBH, the temperature regime of such hypothetical exoplanets corresponds to the habitable zone around main-sequence stars.

Here, presumably for the first time, we  preliminarily examine the effects that general relativity may directly have on the habitability of such putative Earth-like worlds through the long-term de Sitter and Lense--Thirring precessions of their spin axis impacting, among other things, their obliquity, $\varepsilon$. Let us recall that it is defined as the angle between the planet's proper spin angular momentum, $\bds S$, and the orbital angular momentum, $\bds L$ \citep{2016AsBio..16..487B}. Just for illustrative purposes, we will, first, work within the scenario put forth by \citet{2019arXiv191000940S} by keeping $M$ and $r$ fixed. If, on the one hand, such a choice may perhaps be deemed  too restrictive because it picks just a single point in the parameter space, on the other hand, the previous overview should have shown how wide and different are the increasingly numerous scenarios that are being recently  proposed in the literature; it is not possible to treat them here in all their richness. Moreover, even working with one of them like that by \citet{2019arXiv191000940S}, it would be certainly possible, in principle, to vary some of the physical and/or orbital parameters involved, but at the price of investigating less interesting scenarios from a physical point of view as they may refer to uninhabitable planets. Nonetheless, for the sake of completeness, we will investigate also a different SMBH--planet system implying an elliptical path leading the planet from an aponigricon
as large as $Q=200\,R_\mathrm{s}$ to a perinigricon as little as $q=60\,R_\mathrm{s}$, corresponding to an eccentricity $e=0.538$. A scenario with a lighter SMBH with $M=1\times 10^5\,M_\odot$ will be briefly considered as well.

The axial tilt of a planet is a key parameter for the insolation received from the orbited source of electromagnetic waves and thus its ability to sustain life over the eons \citep{1993Natur.361..615L,1997Icar..129..254W,2004A&A...428..261L,2014AsBio..14..277A,2015P&SS..105...43L,
2017arXiv171008052Q,2017ApJ...844..147K,2018AJ....155..237S,2019arXiv191108431Q}. Indeed, variations in obliquity $\Delta\varepsilon(t)$ drive changes in planetary climate. If $\Delta\varepsilon(t)$ are rapid and/or large, the resulting climate shifts can be commensurately severe, as pointed out by, e.g., \citet{2004Icar..171..255A}. In the case of Earth, its obliquity, $\varepsilon_\oplus$, changes slowly with time from $\simeq 22^\circ .1$ to $24^\circ .5$, undergoing an oscillation cycle with $\Delta\varepsilon_\oplus\lesssim 2^\circ. 4$ in about $41,000\,\mathrm{yr}$ \citep{2019arXiv191108431Q}. The value of the Earth's obliquity  impacts the seasonal cycles and its long-term variation affects the terrestrial climate
\citep{Milan1941}, as deduced from geologic records \citep{Kerr1987,1995GeoJI.121...21M,1999E&PSL.174..155P}. Such a moderate and benign modulation of $\varepsilon_\oplus$ is sensibly due to the Moon \citep{1993Natur.361..615L}, where the change would have been larger for a Moonless Earth \citep{1993Natur.361..615L,2012Icar..217...77L,2014ApJ...790...69L}. On the other hand, the obliquity of Mars, $\varepsilon_{\mars}$, whose axis is not strongly stabilized by its tiny moons, experienced variations that have reached $\Delta\varepsilon_{\mars}\simeq 60^\circ$ \citep{1973Sci...181..260W,1993Sci...259.1294T}, and they contributed to the martian atmospheric collapse \citep{2003Natur.426..797H,2005Natur.434..346H,2013Icar..222...81F}, in addition to processes that altered the atmospheric pressure as well
\citep{2018JGRE..123..794M,2019SSRv..215...10K}. Venus, which perhaps was habitable in the early past \citep{2016GeoRL..43.8376W}, may have experienced unexpected long-term variability of up to $\Delta\varepsilon_{\venus}\simeq \pm 7^\circ$ for certain initial values of retrograde rotation, i.e. for $\varepsilon^0_{\venus}>90^\circ$. \citep{2016AsBio..16..487B}.

We will show that, in fact, under certain circumstances, also the general relativistic spin precessions do have potentially an impact on the ability of a rogue planet orbiting a Kerr SMBH, assumed maximally spinning, to sustain life, due to the induced variations of its obliquity. Clearly, also other physical torques of classical nature due to $N$-body interactions with the planet's equatorial bulge might impact $\varepsilon$ and also other orbital parameters, like, e.g.,  $e$, which are relevant for life; however, they depend on specific circumstances: whether the planet orbits a star,\footnote{The third-body tidal effects of the SMBH on the orbital eccentricity, $e_\star$, of the planet's motion around its parent star  and on the obliquity, $\varepsilon_\star$, to its orbital plane were treated by \citet{2020ApJ...889..152I}.}  the presence of one or more large moons, other planets, passing stars, a more or less pronounced equatorial flattening, the interaction with the SMBH's accretion disk itself, etc; their calculation are beyond the scopes of the present paper. Instead, the potentially nonnegligible Einsteinian effects treated here occur simply because the planet moves in the deformed spacetime of the SMBH, irrespectively of any other particular details on its physical characteristics and the presence of other interacting bodies or not. As such, they should enter the overall budget of the possible phenomena treated so far, affecting the habitability of a SMBH's planet, even at in a regime of comparatively weak gravity, i.e. far from the event horizon itself.

For the sake of completeness, we will look also at the precessional angle $\phi$ of the planet's spin axis. Indeed, if on the one hand, precession may not directly affect habitability as it only changes the direction of the spin axis, on the other hand, it tends to accentuate variations owing to other orbital variations \citep{libro17}.

The paper is organized as follows. In Section\,\ref{solter}, we numerically integrate the time-dependent spin evolution of a rogue planet orbiting a maximally spinning Kerr SMBH in both the Sun--Earth type scenario envisaged by \citet{2019arXiv191000940S}, and an another one characterized by large orbital eccentricity leading the planet as close as $60\,R_\mathrm{s}$ to the SMBH (Section\,\ref{altri}). In Section\,\ref{altri}, we will cursorily look at a case with a lighter SMBH as well.
Section\,\ref{fine} summarizes our findings, and offers our conclusions.
\section{The General Relativistic Shift of the Obliquity of the Planetary Spin Axis}\lb{numero}
\subsection{The Sun--Earth scenario}\lb{solter}
Here, we will first consider in detail the situation envisaged in \citet{2019arXiv191000940S} consisting of a putative (Earth-like) planet orbiting a SMBH with $M_\bullet=1\times 10^8\,M_\odot$ at a distance $r=100\,R_\mathrm{s}$ from it in such a way that the apparent size of the lensed accretion disk is that of the Sun as viewed from the Earth. Although it is not strictly necessary from a mere computational point, in order to give greater weight to the considered scenario, it is tacitly assumed that the mass accretion rate of the SMBH is suitably tuned in such a way that the apparent blackbody temperature of the accretion disk is equal to the solar one.

\begin{figure}[H]
\centering
\centerline{
\vbox{
\begin{tabular}{cc}
\epsfxsize= 7.8 cm\epsfbox{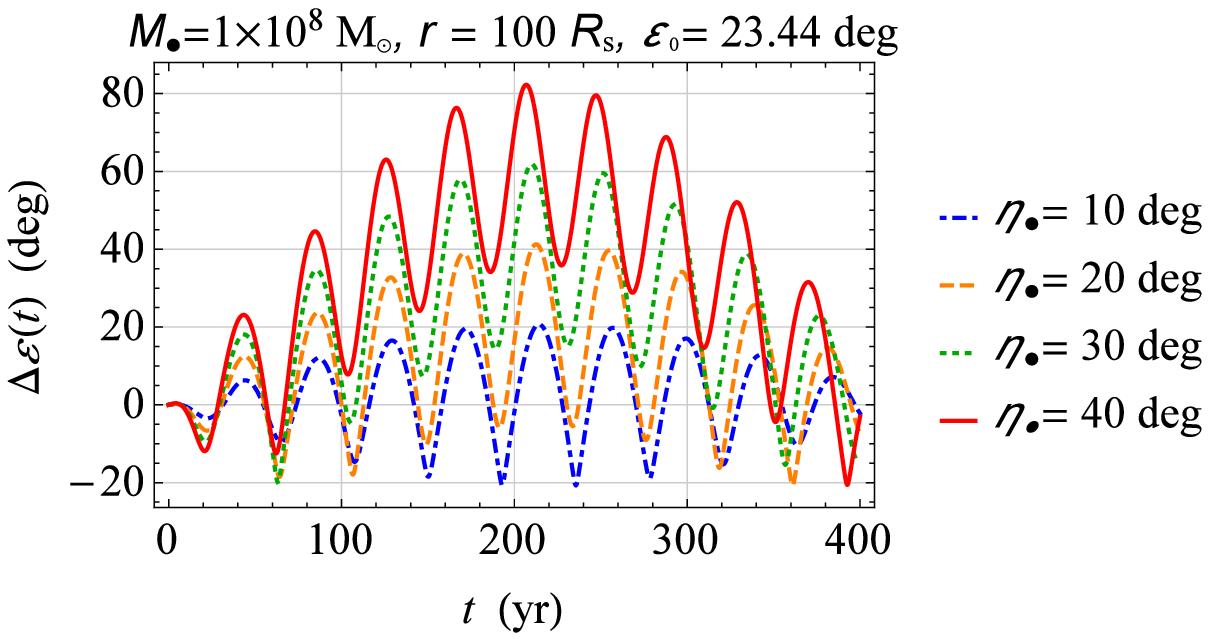} & \epsfxsize= 7.8 cm\epsfbox{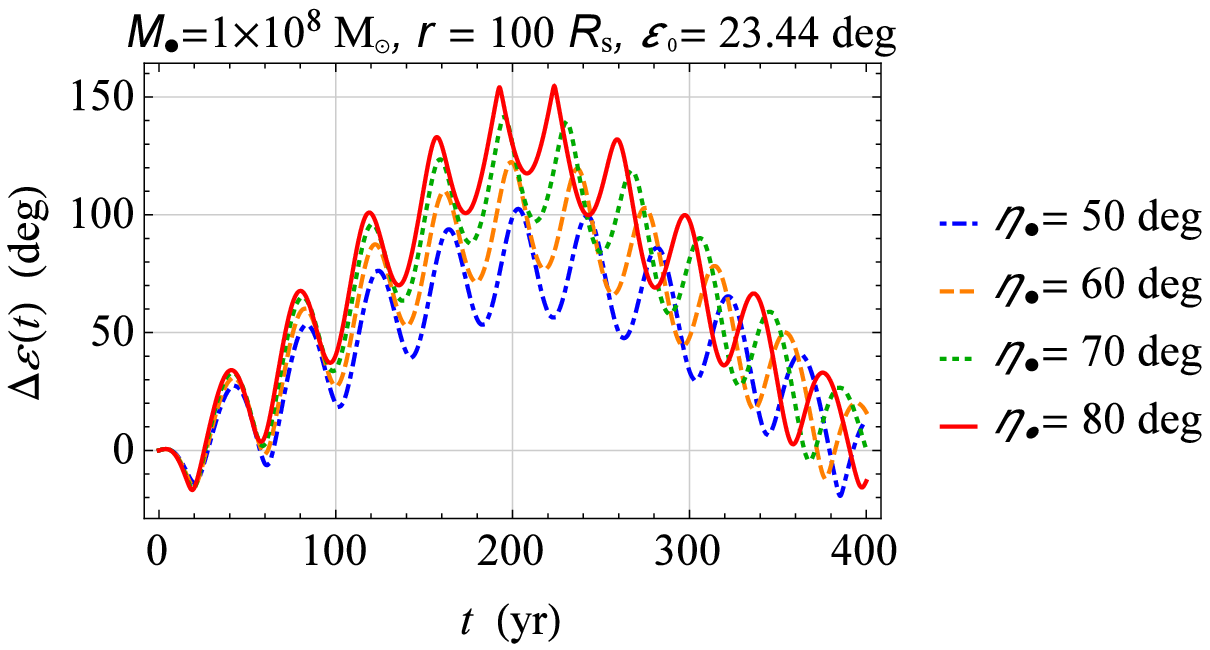}\\
\epsfxsize= 7.8 cm\epsfbox{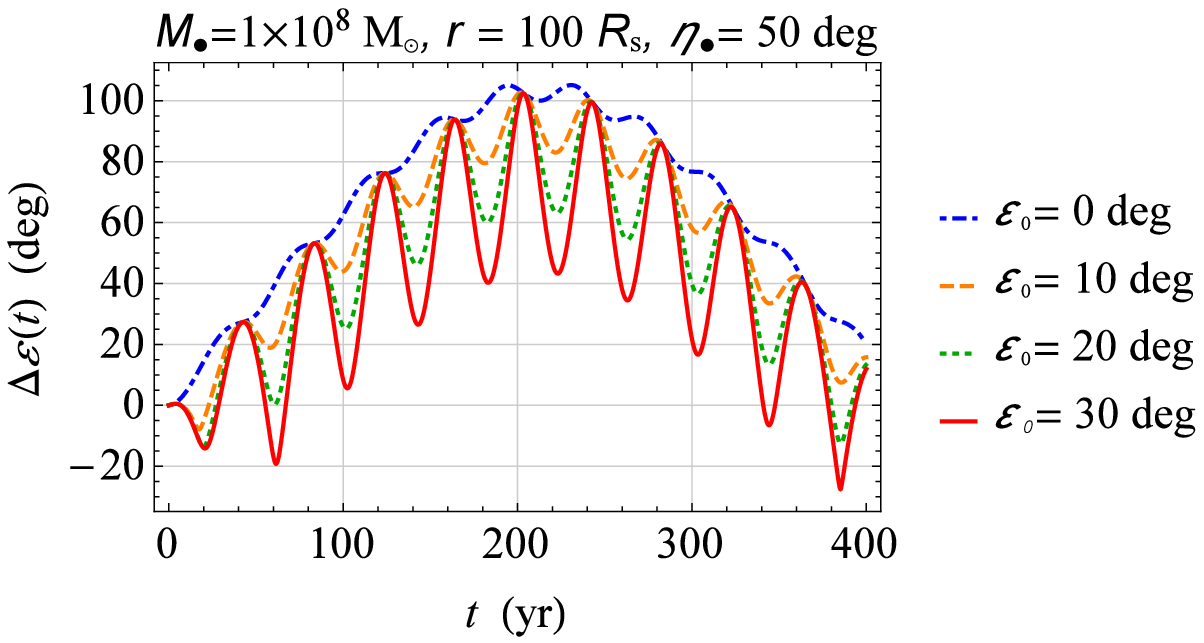} & \epsfxsize= 7.8 cm\epsfbox{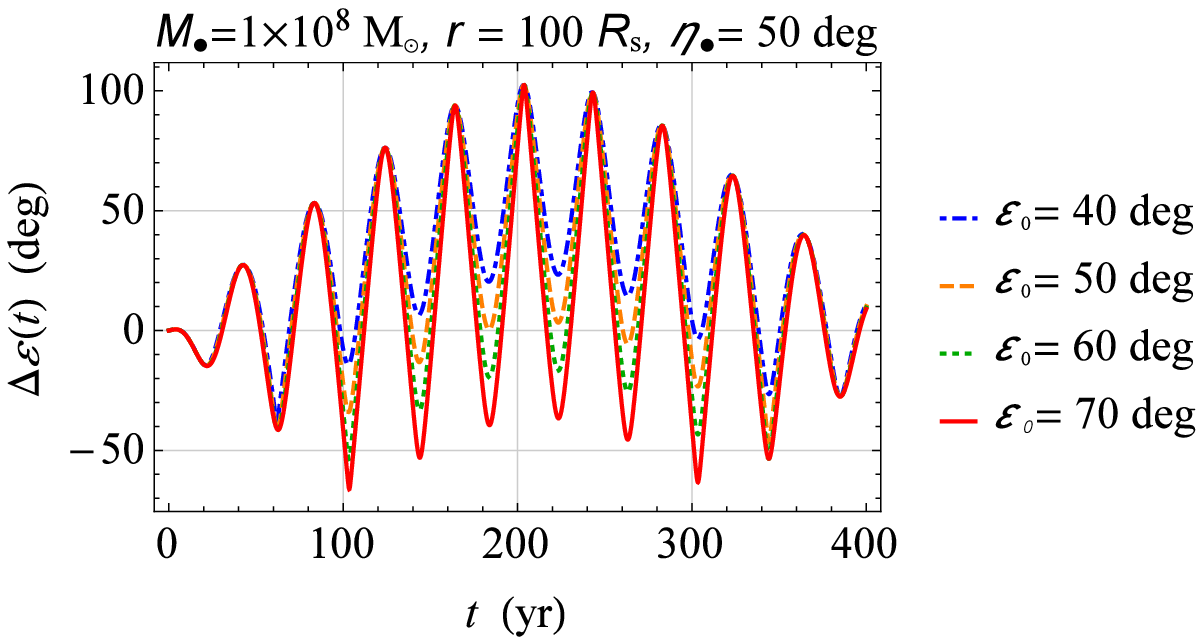}\\
\epsfxsize= 7.8 cm\epsfbox{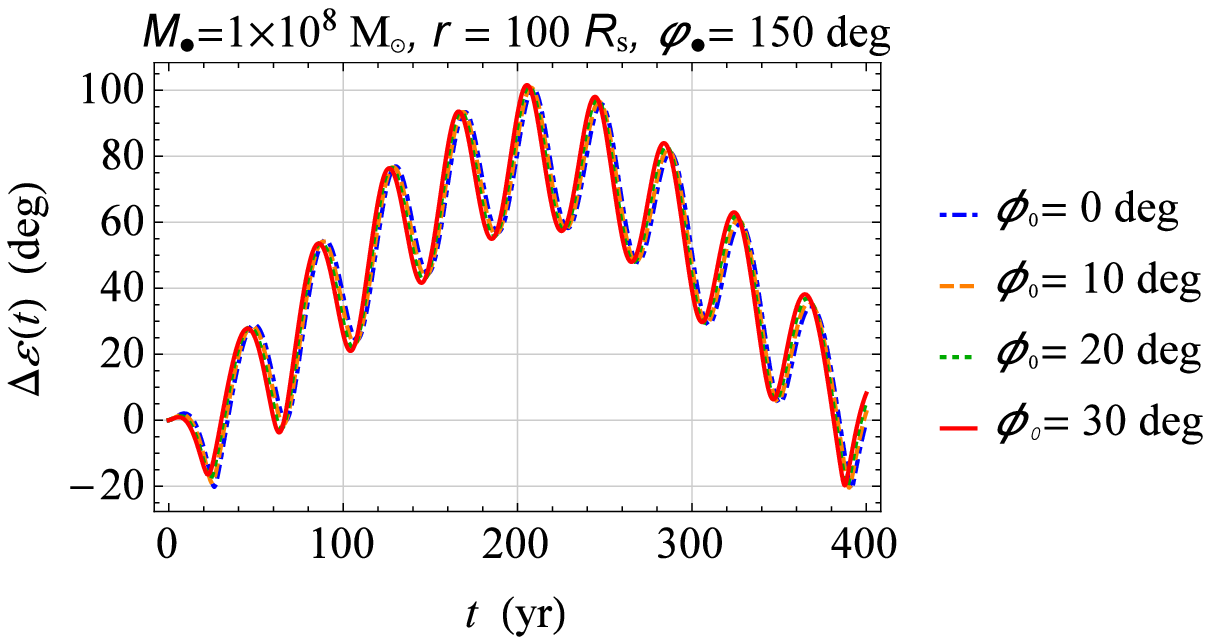} & \epsfxsize= 7.8 cm\epsfbox{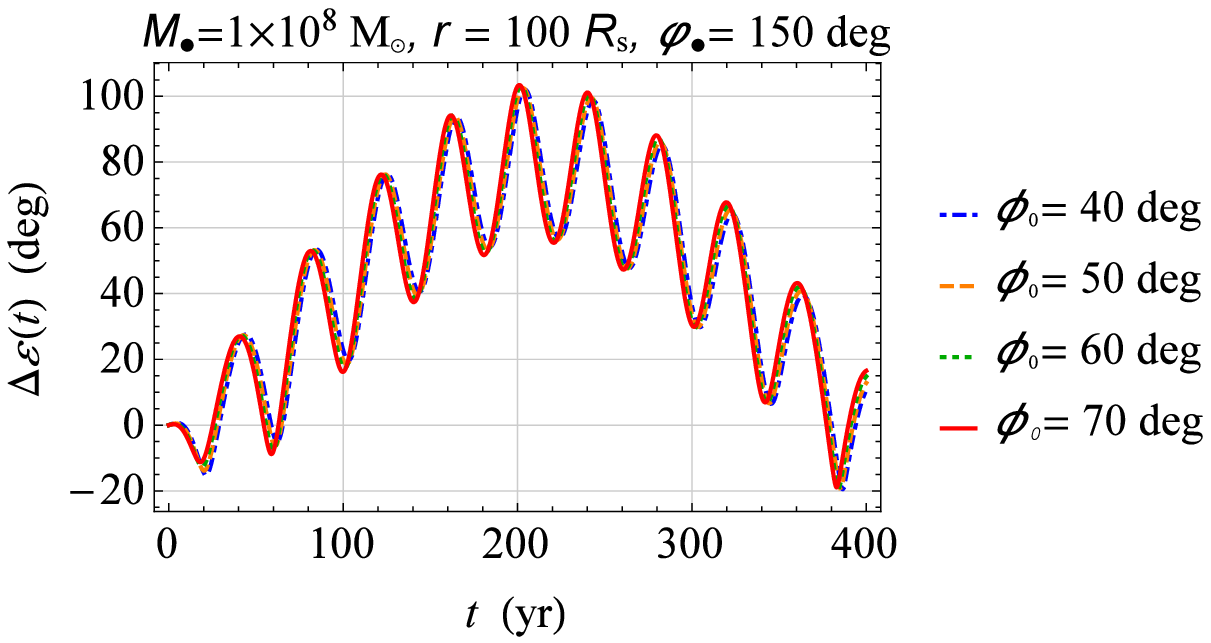}\\
\epsfxsize= 7.8 cm\epsfbox{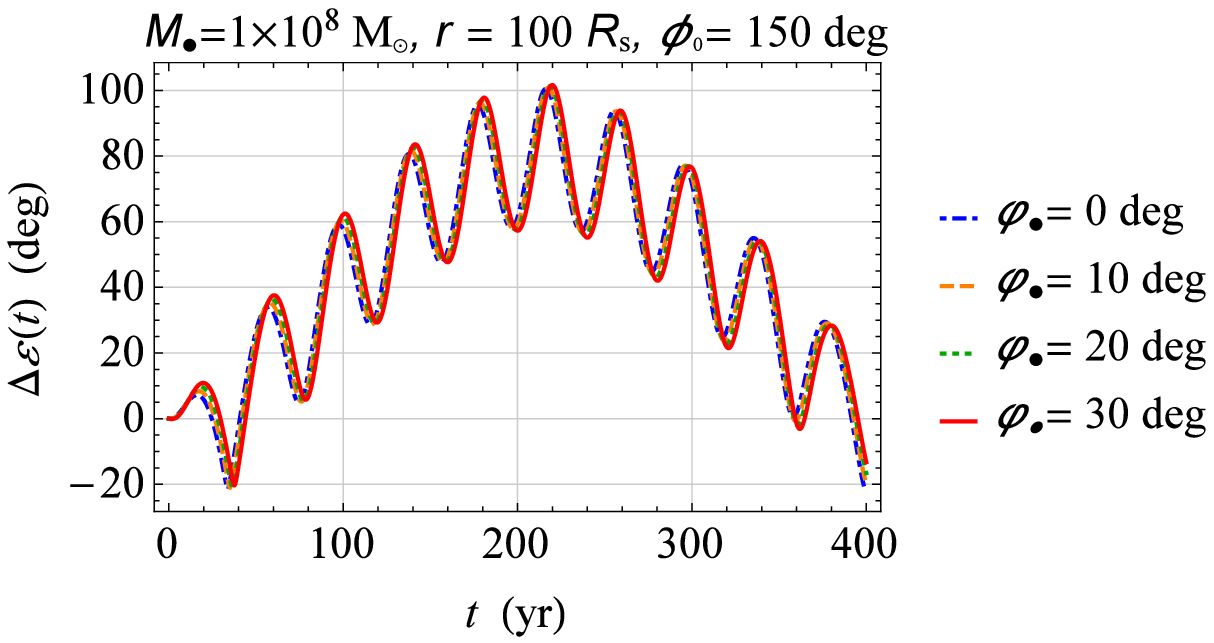} & \epsfxsize= 7.8 cm\epsfbox{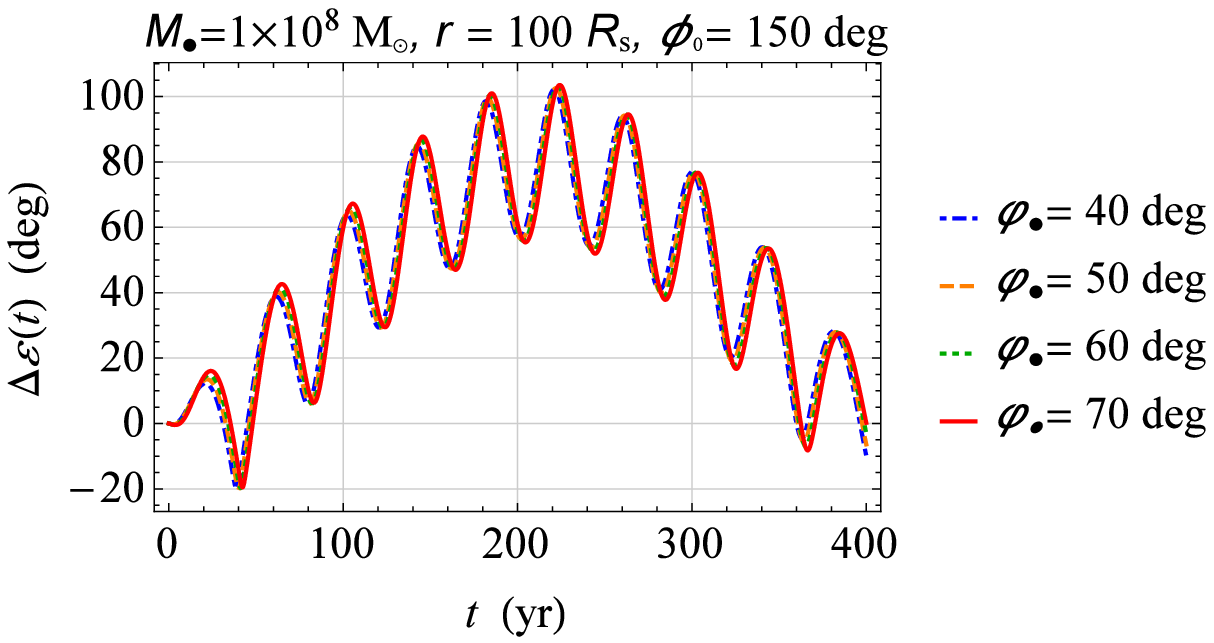}\\
\end{tabular}
}
}
\caption{
Numerically integrated time series of the shift $\Delta\varepsilon\ton{t}$ of the obliquity $\varepsilon$ of the planet's spin axis to the orbital plane for different values of some parameters by assuming $M_\bullet = 1\times 10^8\,M_\odot,\,r = 100\,R_\mathrm{s}$, and an initially circular orbit $(e_0=0)$ with generic initial conditions.  First and second rows: sensitivity to the obliquities $\eta_\bullet,\,\varepsilon_0$ of the spin axes of the SMBH and of the planet, respectively, by assuming  $\varphi_\bullet=150^\circ,\,\phi_0=150^\circ$ for their azimuthal angles. Third and fourth rows: sensitivity to the azimuthal angles $\varphi_\bullet,\,\phi_0$ of the spin axes of the SMBH and of the planet, respectively, by assuming $\eta_\bullet = 50^\circ,\,\varepsilon_0 = 23^\circ .4$ for their obliquities.
  }\label{figura1}
\end{figure}

In a rectangular Cartesian coordinate system centered at the SMBH, we simultaneously integrated the post-Newtonian\footnote{Such an approximation is adequate in the present case, as $v^2/c^2\simeq 5\times 10^{-3}$ , and the largest pN acceleration experienced by the planet, i.e. the 1pN gravitoelectric one, is just $\simeq 1\%$ of the Newtonian monopole. We included also the smaller 1pN spin and quadrupole, and the 2pN gravitoelectric accelerations in our dynamical model.} (pN) equations of motion of the planet \citep{2014grav.book.....P}
along with the pN evolutionary equations of the planet's spin axis \citep{2014grav.book.....P}
\eqi
\dert{\bds{\hat{S}}}{t} = \bds{\Omega}\bds\times\bds{\hat{S}} \lb{dSdt}.
\eqf
In \rfr{dSdt}, the angular velocity vector $\bds\Omega$ is the sum of the pN gravitoelectric de Sitter (dS) and gravitomagnetic Lense--Thirring (LT) terms
\eqi
\bds\Omega \lb{bigO}= {\bds\Omega}_\mathrm{dS} + {\bds\Omega}_\mathrm{LT},
\eqf
with \citep{2014grav.book.....P}
\begin{align}
{\bds\Omega}_\mathrm{dS} \lb{OdS}& = \rp{3\,GM_\bullet}{2\, c^2\, r^2}\bds{\hat{r}}\bds\times\bds{v}, \\ \nonumber \\
{\bds\Omega}_\mathrm{LT} \lb{OLT}& = \rp{G\,J^\bullet}{2\, c^2\, r^3}\qua{3\ton{{\bds{\hat{J}}}^\bullet\bds\cdot\bds{\hat{r}}}\bds{\hat{r}} -{\bds{\hat{J}}}^\bullet  }.
\end{align}
In \rfrs{OdS}{OLT}, $\bds r,\,\bds v$ are the planet's position and velocity, respectively, with respect to the SMBH, assumed maximally rotating.
In the numerical integration, the \virg{ecliptic} plane, i.e. the orbital plane itself, was assumed as reference $\grf{x,\,y}$ plane so that the inclination is $I=0$. Furthermore, the spin axes of the planet and of the SMBH $\bds{\hat{S}},\,{\bds{\hat{J}}}^\bullet$ were parameterized as
\begin{align}
\hat{S}_x \lb{Sx}& =\sin\varepsilon\,\cos\phi, \\ \nonumber \\
\hat{S}_y & =\sin\varepsilon\,\sin\phi, \\ \nonumber \\
\hat{S}_z \lb{Sz} & =\cos\varepsilon, \\ \nonumber \\
\hat{J}^\bullet_x & =\sin\eta_\bullet\,\cos\varphi_\bullet, \\ \nonumber \\
\hat{J}^\bullet_y & =\sin\eta_\bullet\,\sin\varphi_\bullet, \\ \nonumber \\
\hat{J}^\bullet_z \lb{Jz}& =\cos\eta_\bullet,
\end{align}
where $\varepsilon,\,\eta_\bullet$ are the obliquities of  $\bds{\hat{S}},\,{\bds{\hat{J}}}^\bullet$, respectively, to the orbital plane;  $\phi,\,\varphi_\bullet$ are their azimuthal angles.
Fixed generic initial conditions were adopted for the planet's orbital motion around the SMBH, assumed initially circular\footnote{Here, $e_0$ denotes the initial value of the orbital eccentricity $e$.} $\ton{e_0=0}$, while the initial orientations of the spin axes were varied from one run to another spanning a time interval $\Delta t = 400\,\mathrm{yr}$ long. Figure\,\ref{figura1} depicts our results for the change $\Delta\varepsilon\ton{t}$ of the planet's obliquity experienced with respect to its initial value $\varepsilon_0$.
\begin{figure}[H]
\centering
\centerline{
\vbox{
\begin{tabular}{cc}
\epsfxsize= 7.8 cm\epsfbox{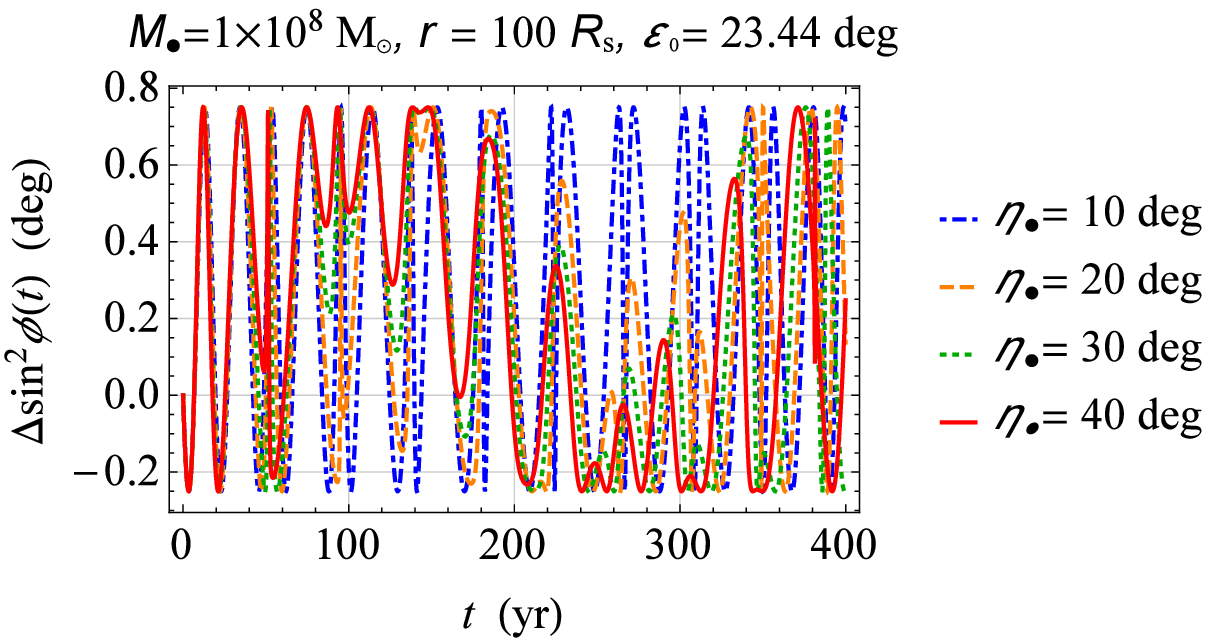} & \epsfxsize= 7.8 cm\epsfbox{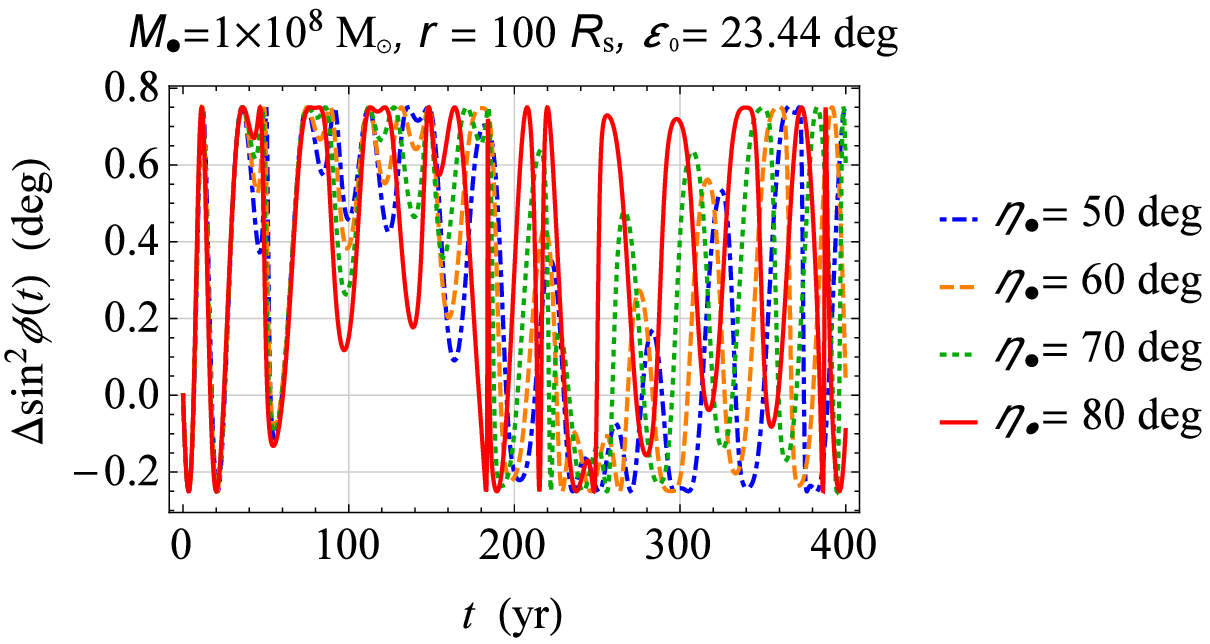}\\
\epsfxsize= 7.8 cm\epsfbox{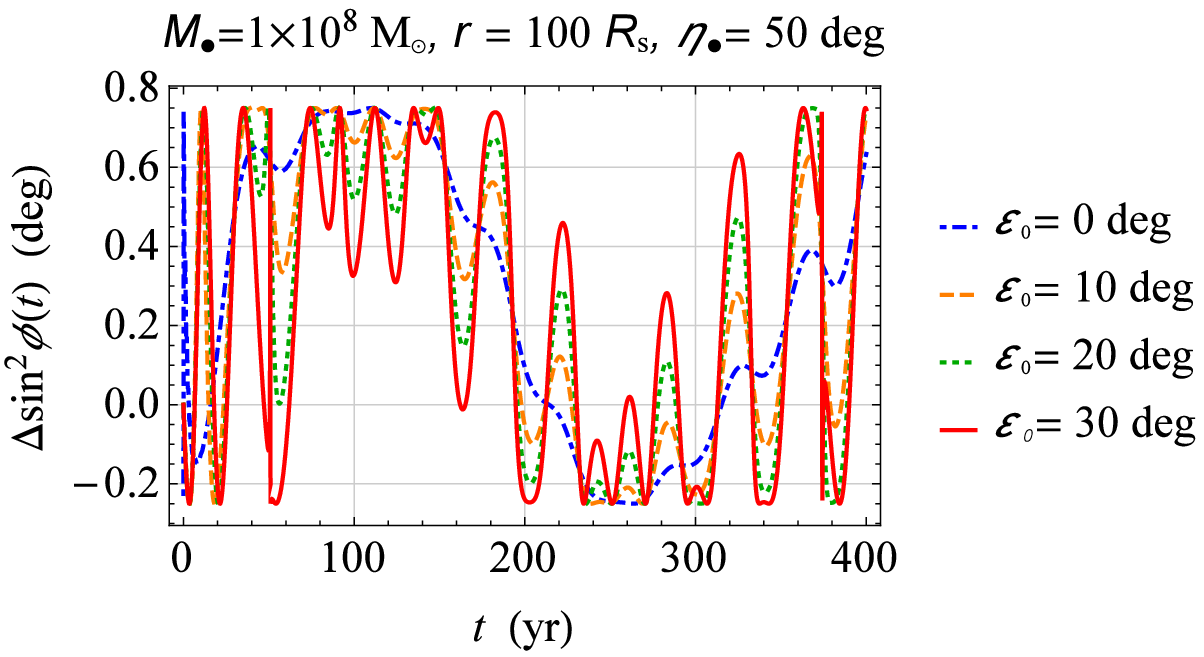} & \epsfxsize= 7.8 cm\epsfbox{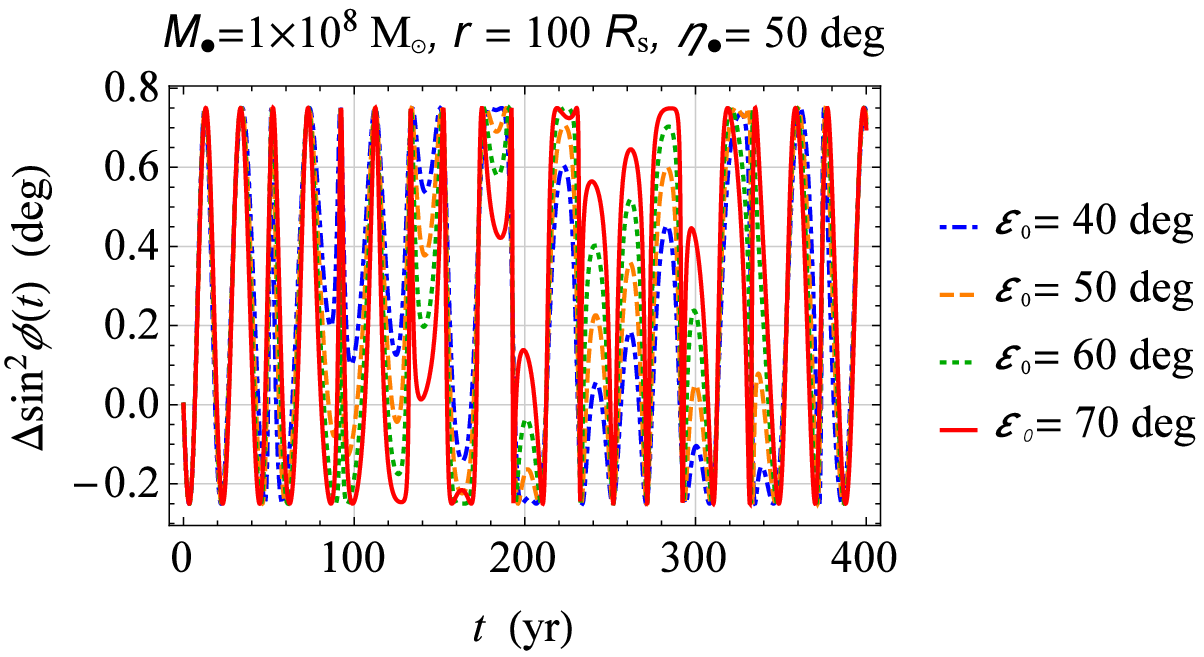}\\
\epsfxsize= 7.8 cm\epsfbox{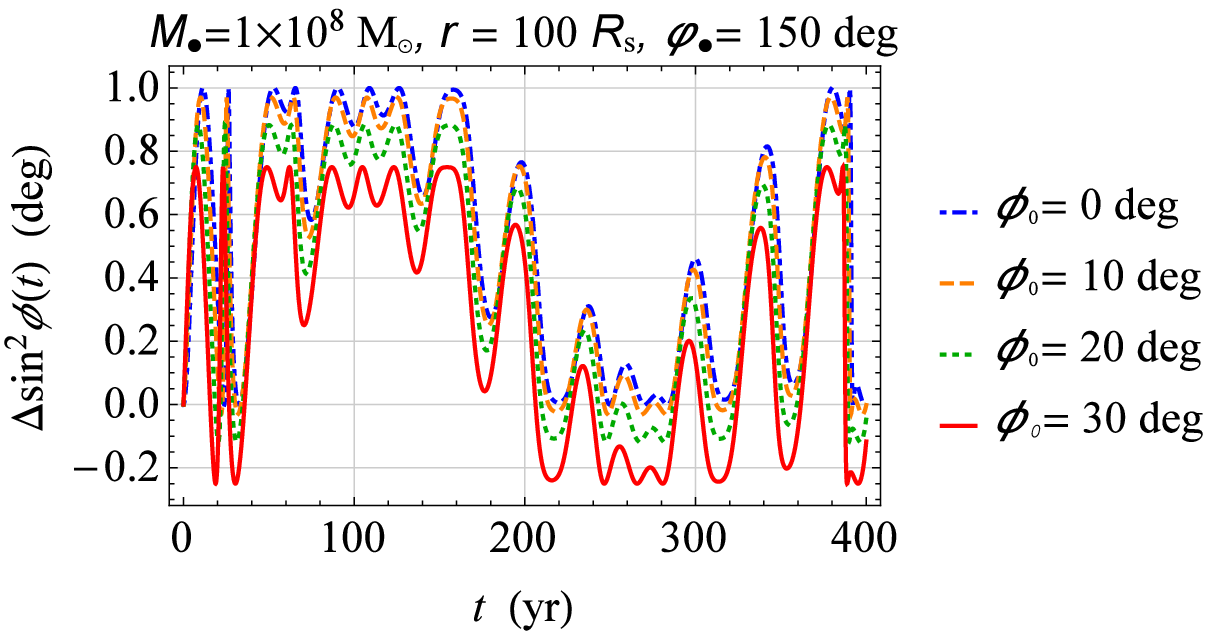} & \epsfxsize= 7.8 cm\epsfbox{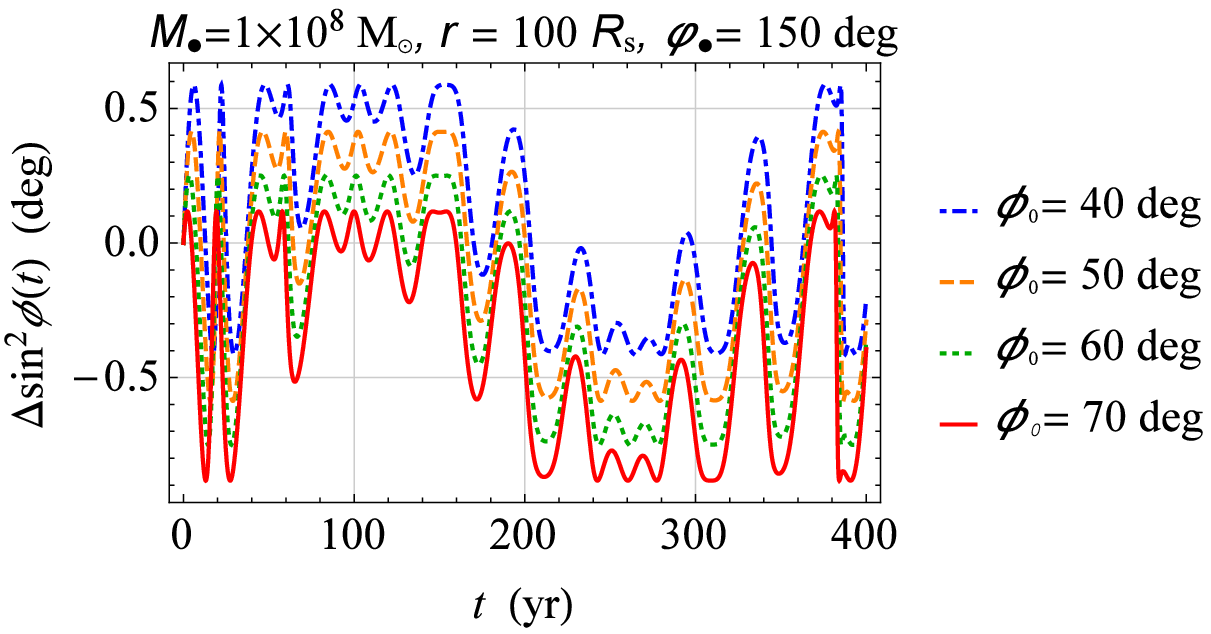}\\
\epsfxsize= 7.8 cm\epsfbox{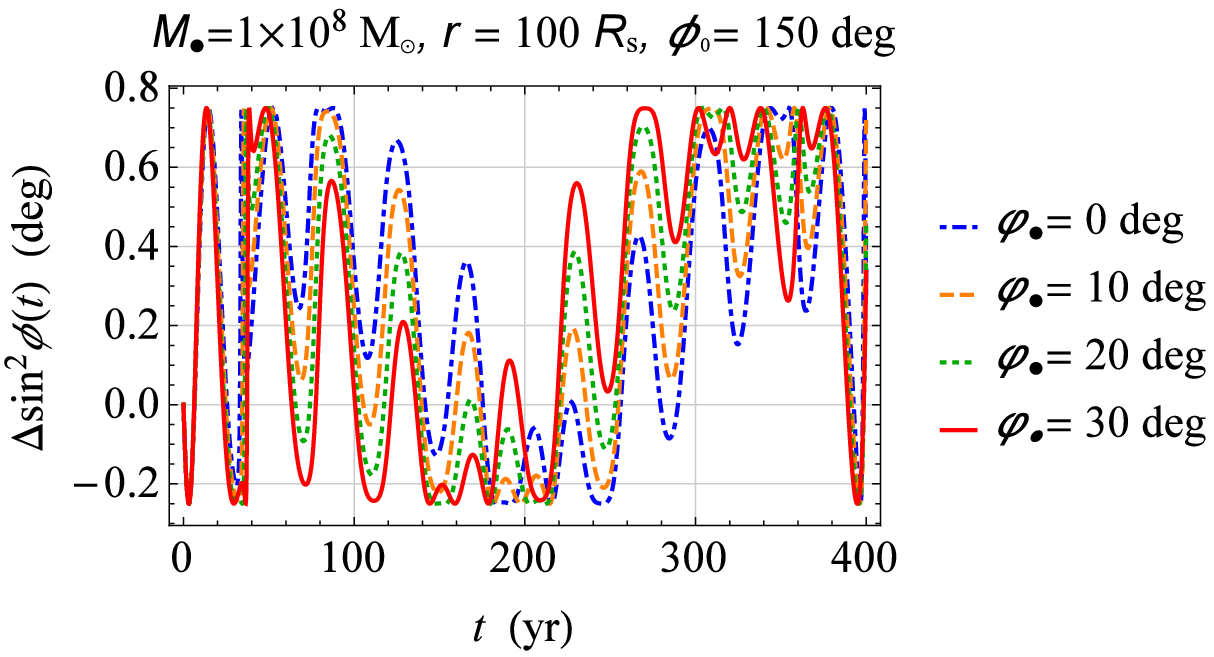} & \epsfxsize= 7.8 cm\epsfbox{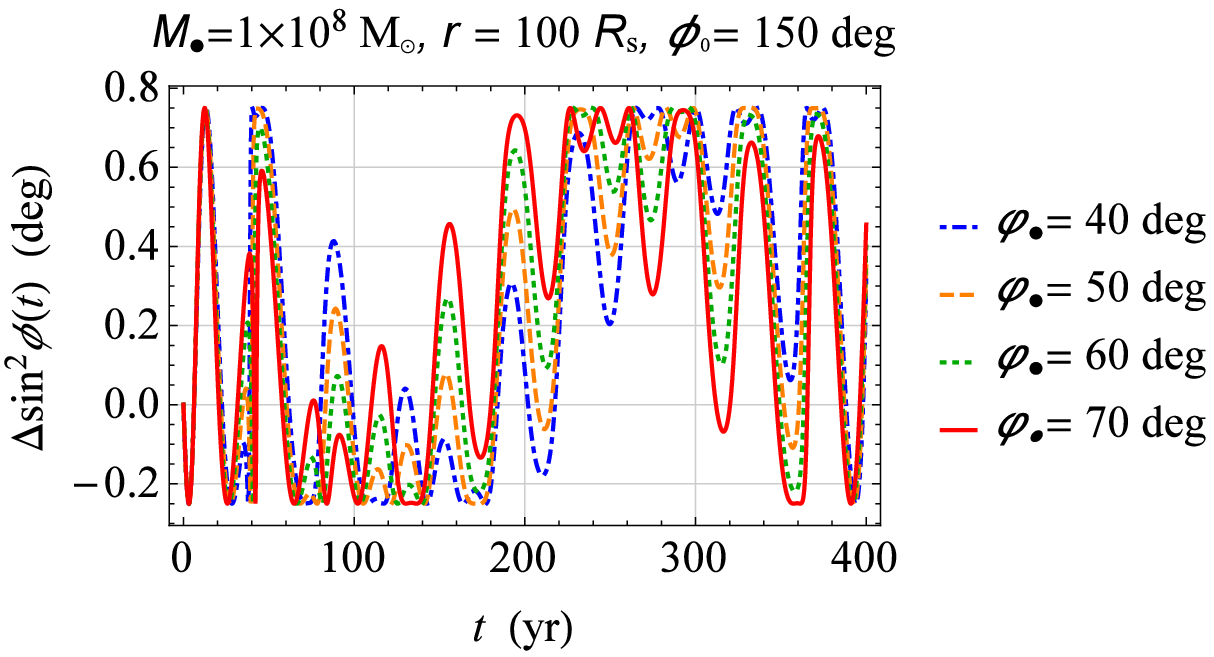}\\
\end{tabular}
}
}
\caption{
Numerically integrated time series of the shift $\Delta\sin^2\phi\ton{t}$, where $\phi(t)$ is the precession angle of the planet's spin axis, for different values of some parameters by assuming $M_\bullet = 1\times 10^8\,M_\odot,\,r = 100\,R_\mathrm{s}$, and an initially circular orbit $(e_0=0)$ with generic initial conditions.  First and second rows: sensitivity to the obliquities $\eta_\bullet,\,\varepsilon_0$ of the spin axes of the SMBH and of the planet, respectively, by assuming  $\varphi_\bullet=150^\circ,\,\phi_0=150^\circ$ for their azimuthal angles. Third and fourth rows: sensitivity to the azimuthal angles $\varphi_\bullet,\,\phi_0$ of the spin axes of the SMBH and of the planet, respectively, by assuming $\eta_\bullet = 50^\circ,\,\varepsilon_0 = 23^\circ .4$ for their obliquities.}\label{figura2}
\end{figure}
From the first row of Figure\,\ref{figura1}, it can be noted that the numerically integrated shift $\Delta\varepsilon\ton{t}$ of the planet's spin obliquity, obtained by starting from an initial value $\varepsilon_0=23^\circ .4$ identical to that of the Earth, exhibits rapidly varying changes, and depends strongly on the obliquity of the SMBH's spin axis $\eta_\bullet$. Even for a small inclination of the orbital plane to the SMBH's equator $(\eta_\bullet \simeq 10^\circ-20^\circ)$, the resulting variation of the tilting of the planet's equator can reach values as large as $\left|\Delta\varepsilon\right|\lesssim 20^\circ-40^\circ$. They can even become as large as  $\left|\Delta\varepsilon\right|\lesssim 80^\circ-150^\circ$ for $\eta_\bullet$ approaching $\simeq 90^\circ$.
The second row of Figure\,\ref{figura1} shows that also the initial value $\varepsilon_0$ of the tilt $\varepsilon\ton{t}$ of the planet's equator to the orbital plane has a certain influence on its time evolution, although less marked than $\eta_\bullet$. It can be noted by comparing the dashed--dotted blue curve in the upper right panel of Figure\,\ref{figura1}, corresponding to $\varepsilon_0=23^\circ .4,\,\eta_\bullet = 50^\circ$, to the somewhat different $\varepsilon_0$-dependent curves in the second row of Figure\,\ref{figura1} obtained for $\eta_\bullet = 50^\circ$.
According to the third and  fourth rows of Figure\,\ref{figura1}, $\Delta\varepsilon\ton{t}$ is substantially insensitive to the azimuthal angles $\phi_0,\,\varphi_\bullet$ of the spin axes of both the planet and the SMBH.

In Figure\,\ref{figura2}, we display the shifts $\Delta\sin^2\phi\ton{t}\doteq\sin^2\phi\ton{t}-\sin^2\phi_0$ related to the precessional angle $\phi\ton{t}$ of the planet's spin axis. Also in this case, the more marked dependence on $\eta_\bullet,\,\varepsilon_0$ with respect to $\phi_0,\,\varphi_\bullet$ is apparent.

\begin{figure}[H]
\centering
\centerline{
\vbox{
\begin{tabular}{cc}
\epsfxsize= 7.8 cm\epsfbox{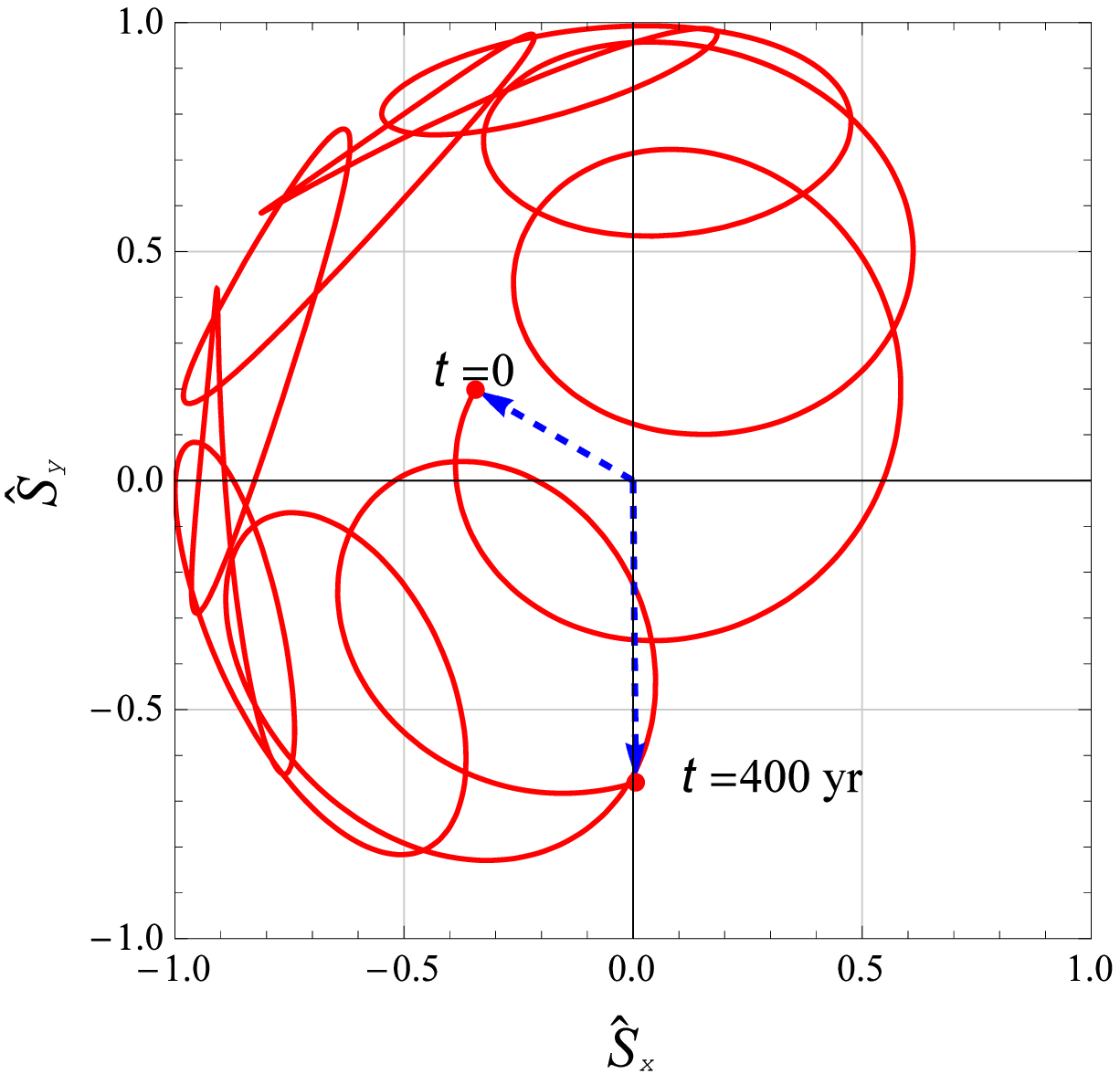} & \epsfxsize= 7.8 cm\epsfbox{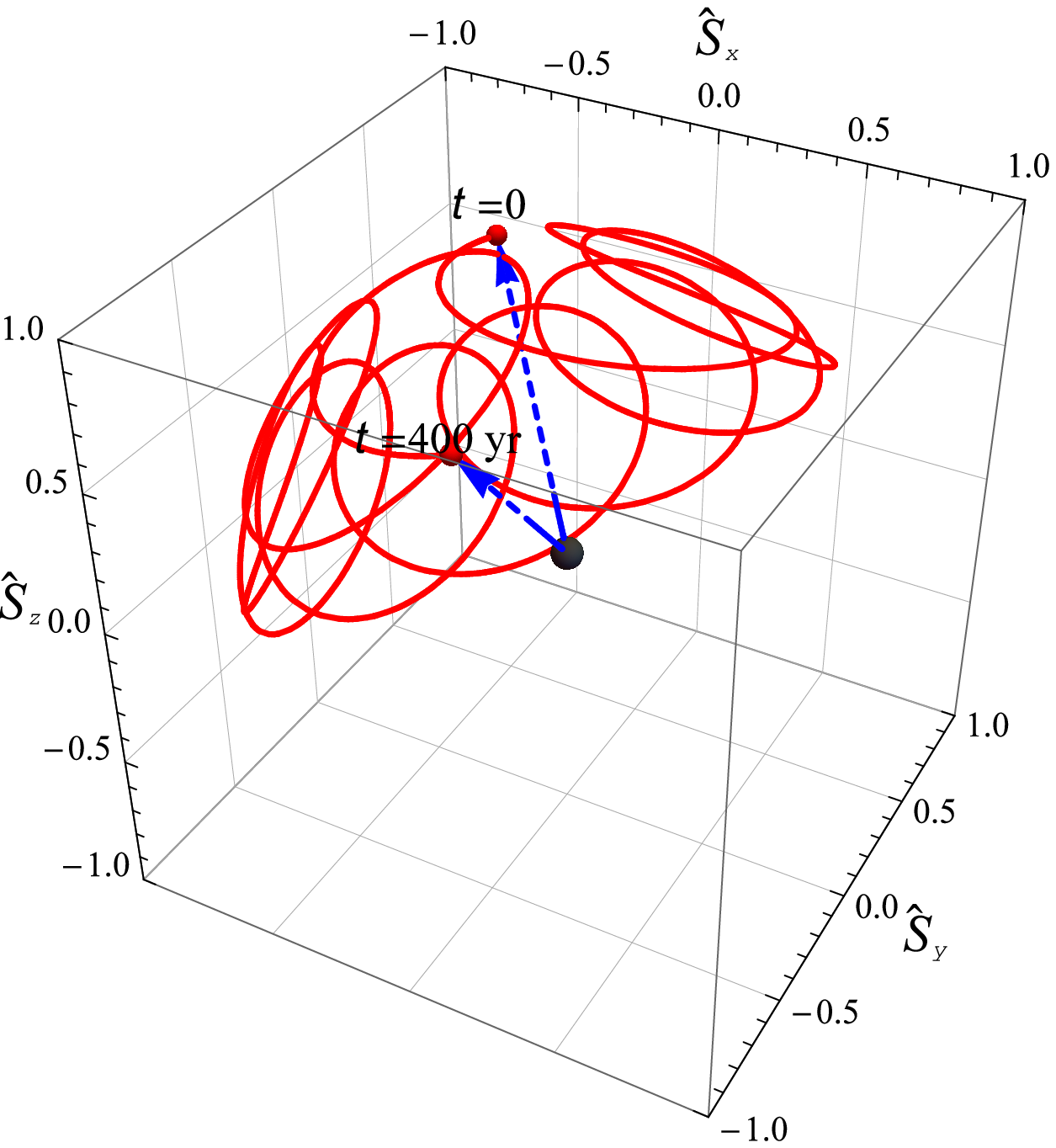}\\
\end{tabular}
}
}
\caption{
Left panel: numerically integrated hodograph, in red, of the planet's spin axis projection, represented by the dashed blue arrow, onto the $\grf{{\hat{S}}_x,\,{\hat{S}}_y}$ plane over $\Delta t=400\,\mathrm{yr}$. Right panel: numerically integrated hodograph, in red, of the planet's spin axis (dashed blue arrow) over the same time span. In both cases, a maximally rotating SMBH with $M_\bullet = 1\times 10^8\,M_\odot,\,\eta_\bullet=50^\circ,\,\varphi_\bullet=150^\circ$ is assumed. The initial conditions of the spin axis of the planet, which orbits at $r=100\,R_\mathrm{s}$ from the SMBH, are $\varepsilon_0=23^\circ .4,\,\phi_0=150^\circ$. Cfr. with the dashed--dotted blue curve in the upper right panel of Figure\,\ref{figura1}.
  }\label{figura3}
\end{figure}

Figure\,\ref{figura3} depicts the spatial path of the planet's spin axis of one of the numerical integrations of the upper right panels of  Figure\,\ref{figura1} and Figure\,\ref{figura2}.
\subsection{Other scenarios}
\begin{figure}[H]
\centering
\centerline{
\vbox{
\begin{tabular}{cc}
\epsfxsize= 7.8 cm\epsfbox{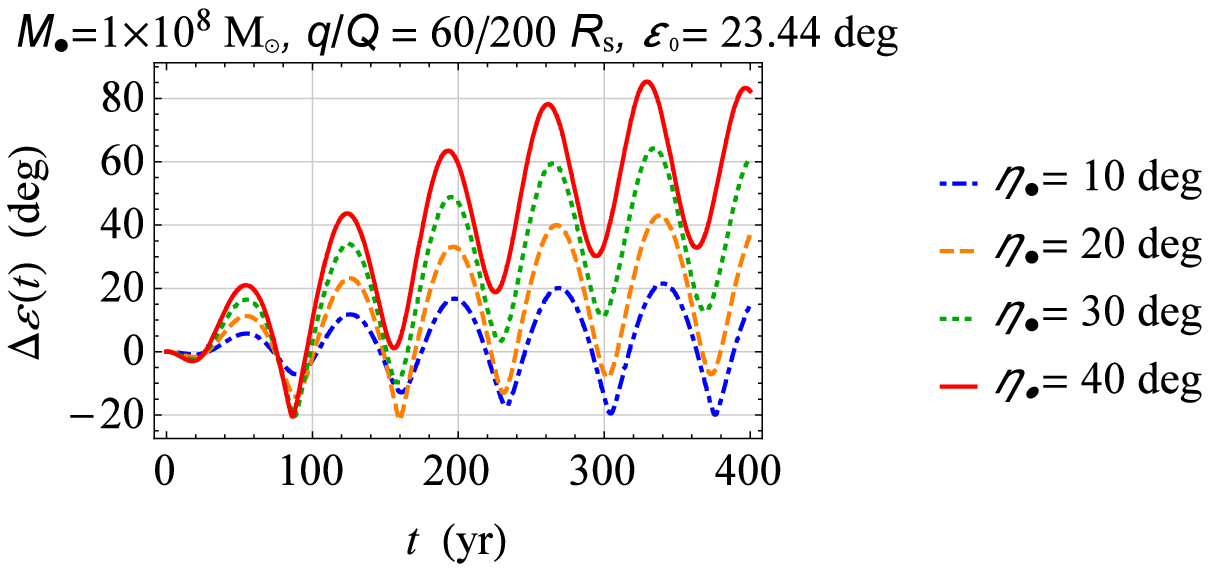} & \epsfxsize= 7.8 cm\epsfbox{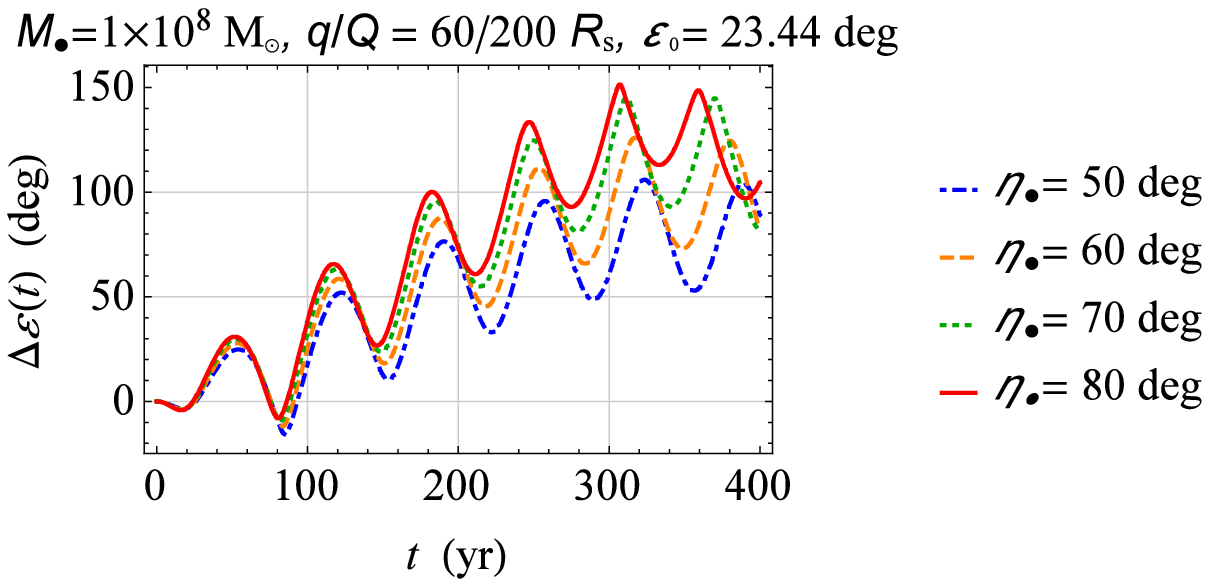}\\
\epsfxsize= 7.8 cm\epsfbox{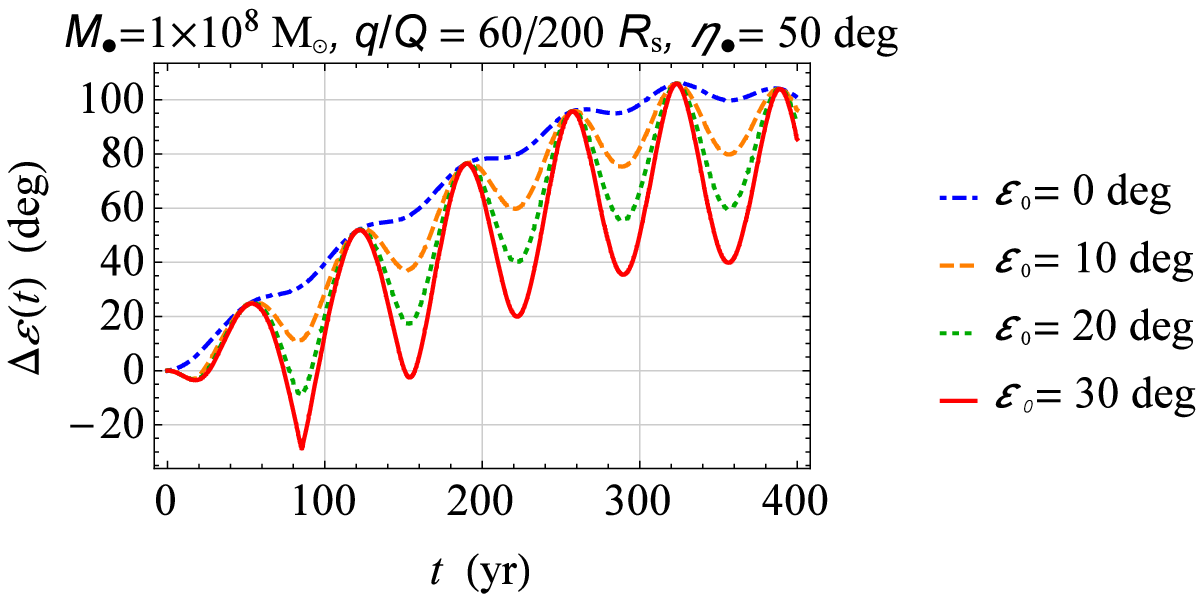} & \epsfxsize= 7.8 cm\epsfbox{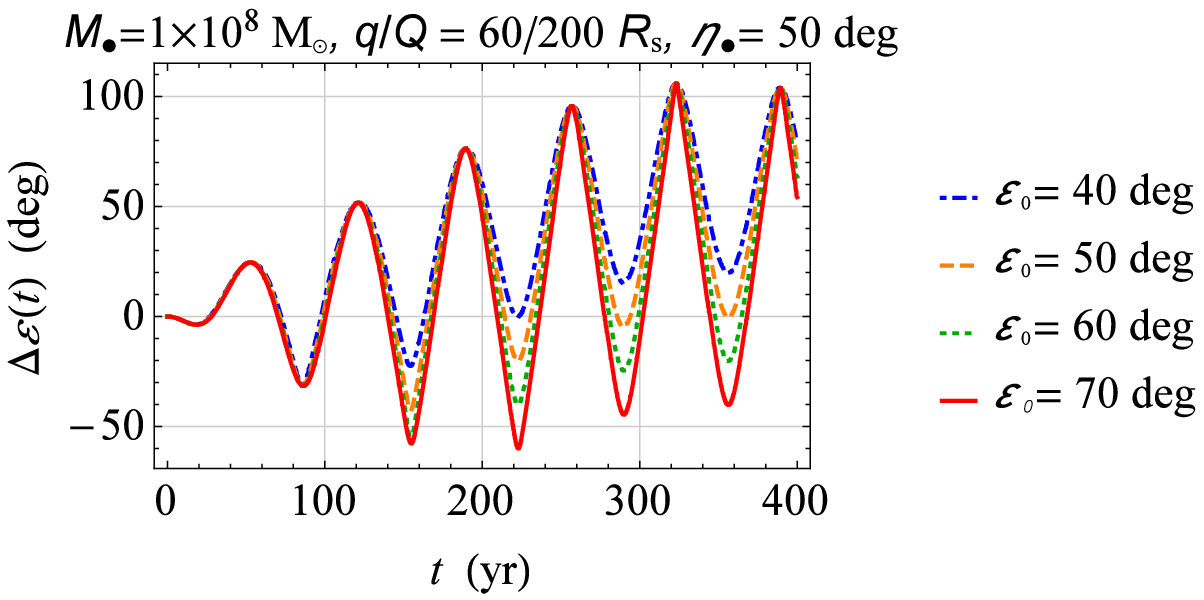}\\
\epsfxsize= 7.8 cm\epsfbox{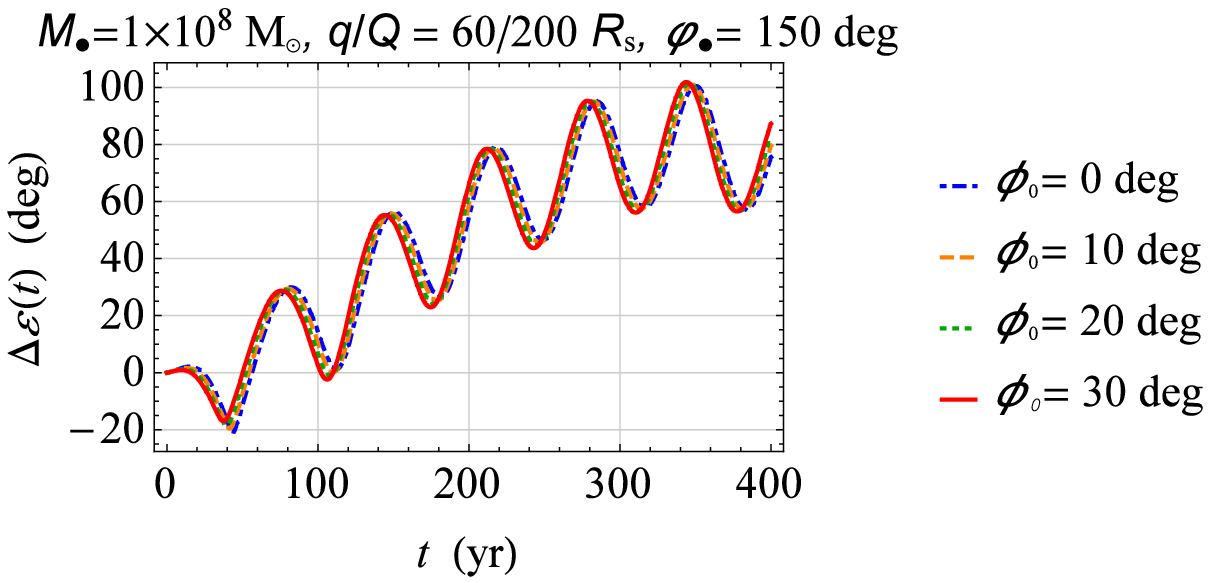} & \epsfxsize= 7.8 cm\epsfbox{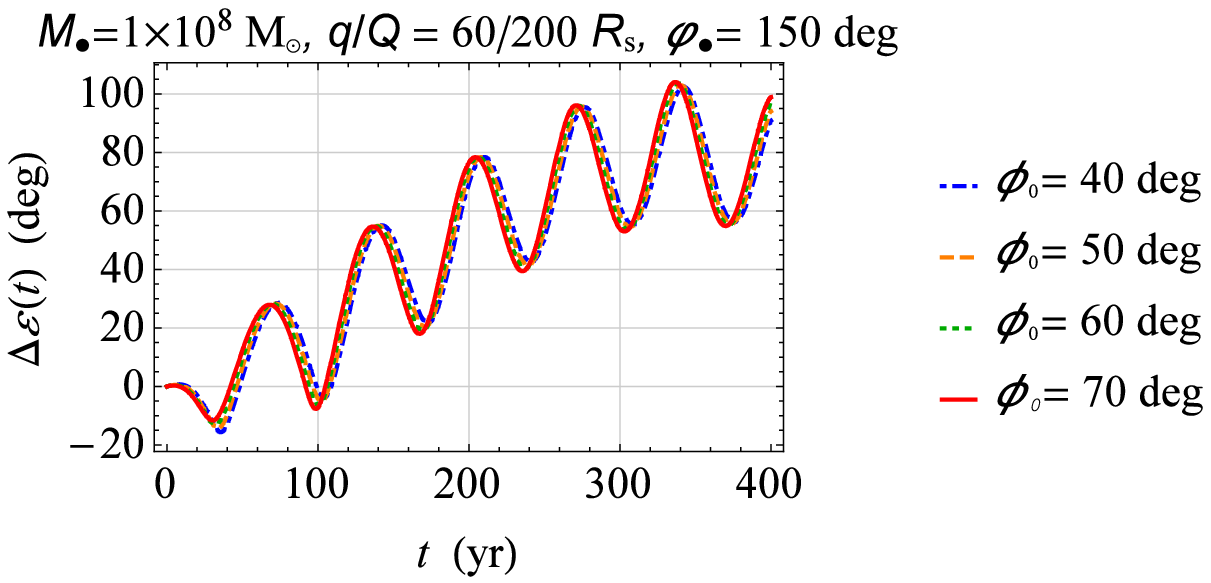}\\
\epsfxsize= 7.8 cm\epsfbox{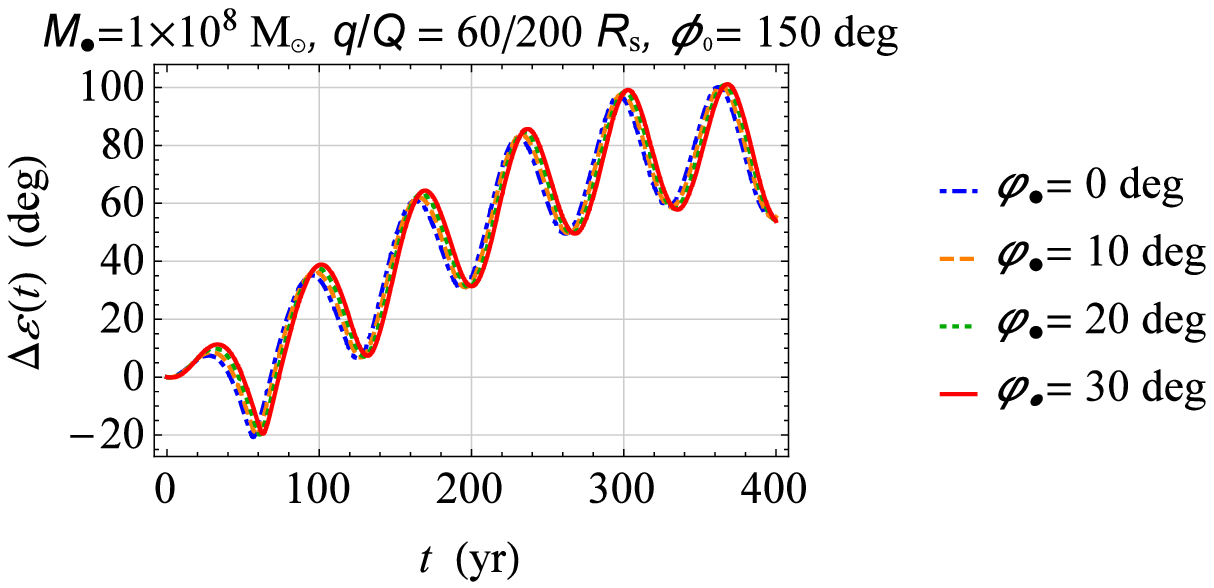} & \epsfxsize= 7.8 cm\epsfbox{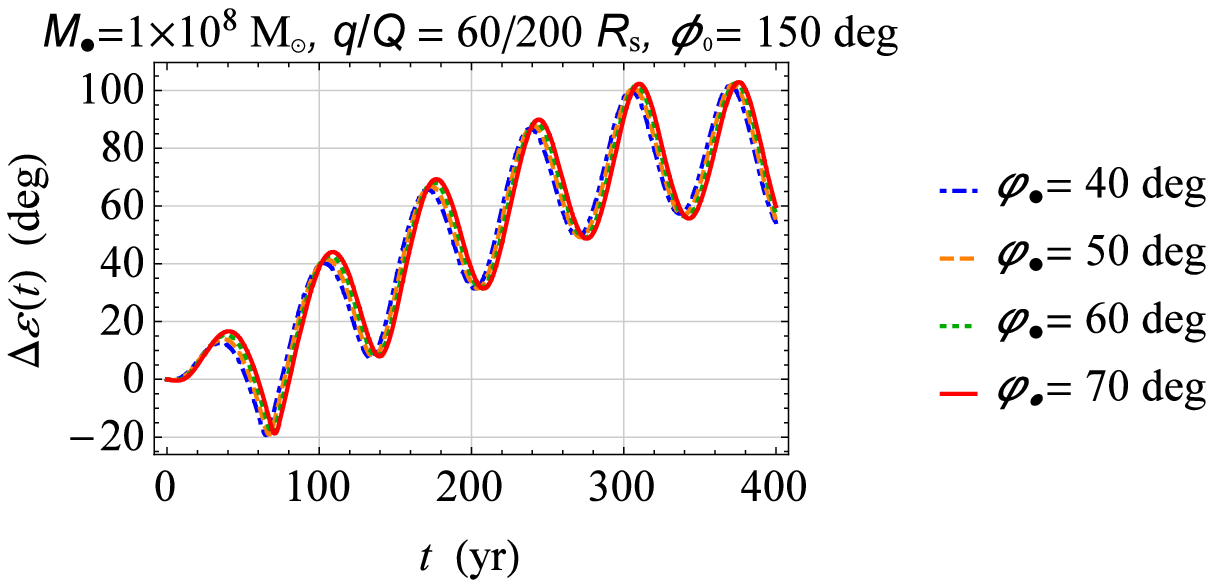}\\
\end{tabular}
}
}
\caption{
Numerically integrated time series of the shift $\Delta\varepsilon\ton{t}$ of the obliquity $\varepsilon$ of the planet's spin axis to the orbital plane for different values of some parameters by assuming $M_\bullet = 1\times 10^8\,M_\odot,\,q = 60\,R_\mathrm{s},\,Q=200\,R_\mathrm{s}$, and generic initial conditions.  First and second rows: sensitivity to the obliquities $\eta_\bullet,\,\varepsilon_0$ of the spin axes of the SMBH and of the planet, respectively, by assuming  $\varphi_\bullet=150^\circ,\,\phi_0=150^\circ$ for their azimuthal angles. Third and fourth rows: sensitivity to the azimuthal angles $\varphi_\bullet,\,\phi_0$ of the spin axes of the SMBH and of the planet, respectively, by assuming $\eta_\bullet = 50^\circ,\,\varepsilon_0 = 23^\circ .4$ for their obliquities.}\label{figura4}
\end{figure}

\begin{figure}[H]
\centering
\centerline{
\vbox{
\begin{tabular}{cc}
\epsfxsize= 7.8 cm\epsfbox{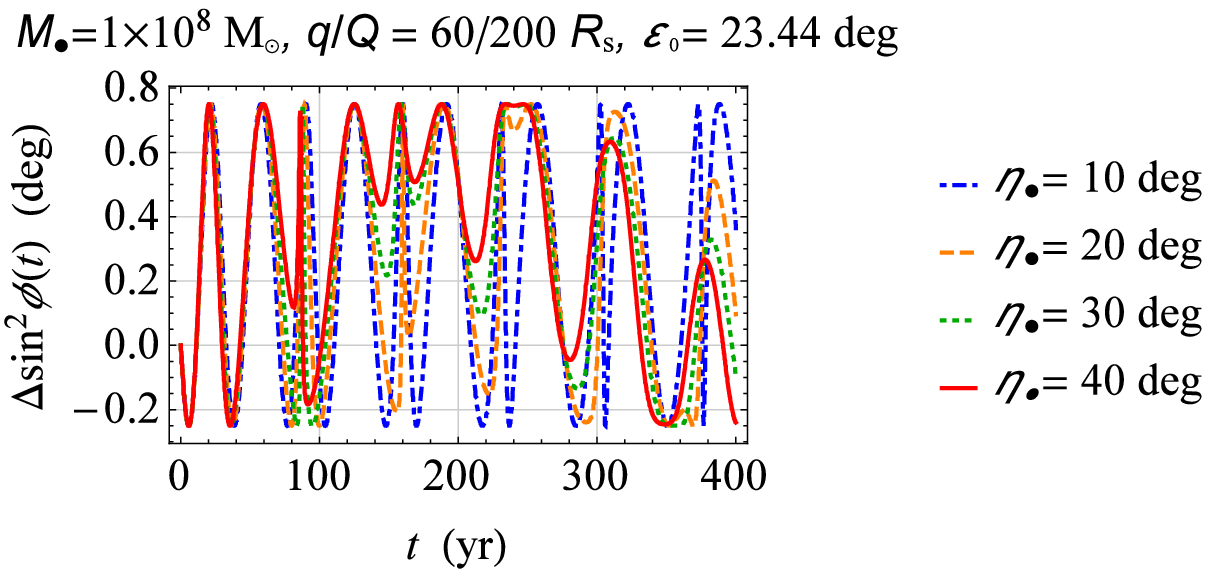} & \epsfxsize= 7.8 cm\epsfbox{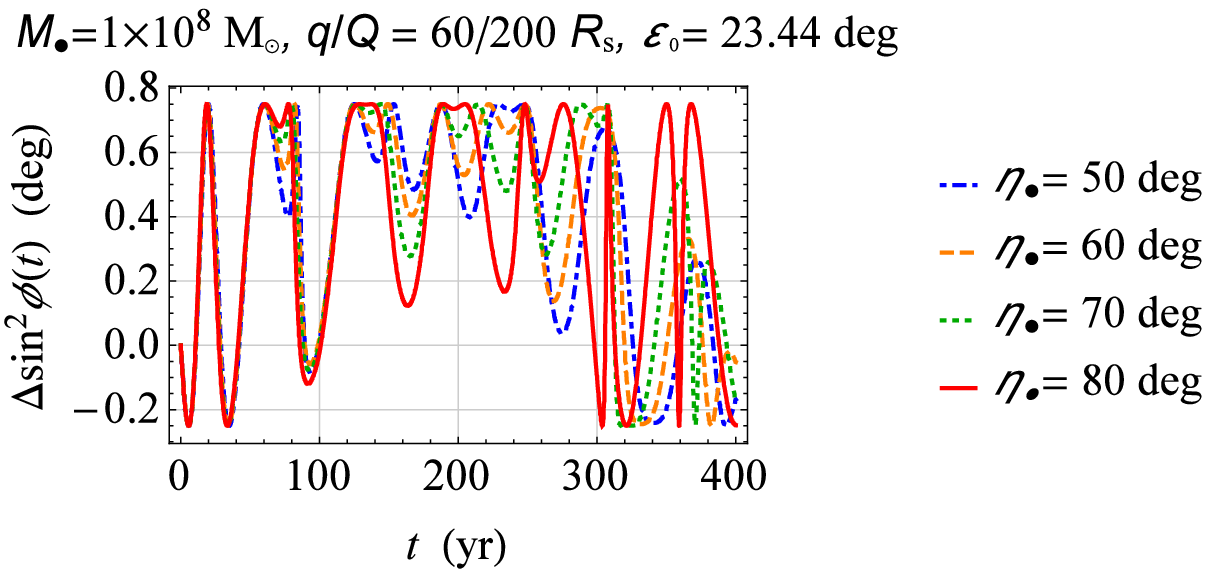}\\
\epsfxsize= 7.8 cm\epsfbox{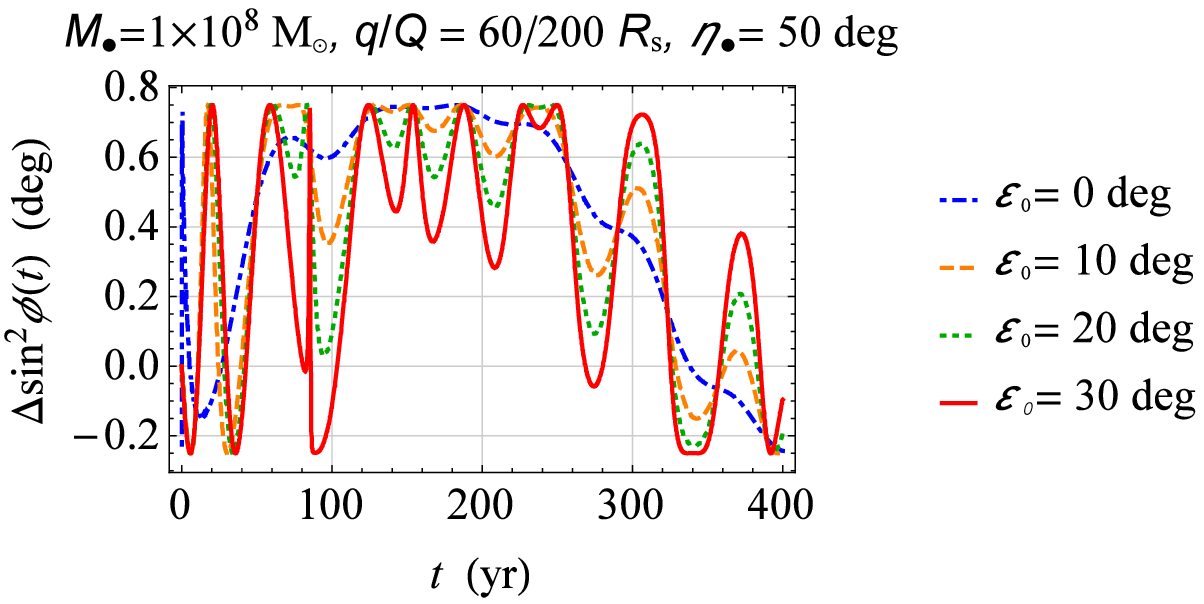} & \epsfxsize= 7.8 cm\epsfbox{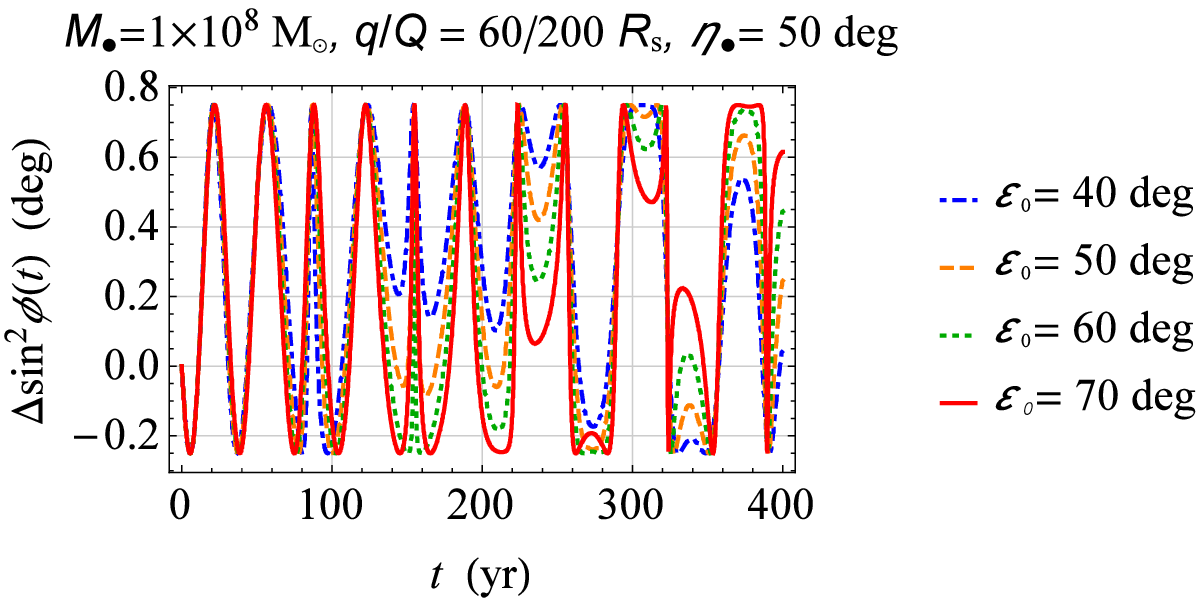}\\
\epsfxsize= 7.8 cm\epsfbox{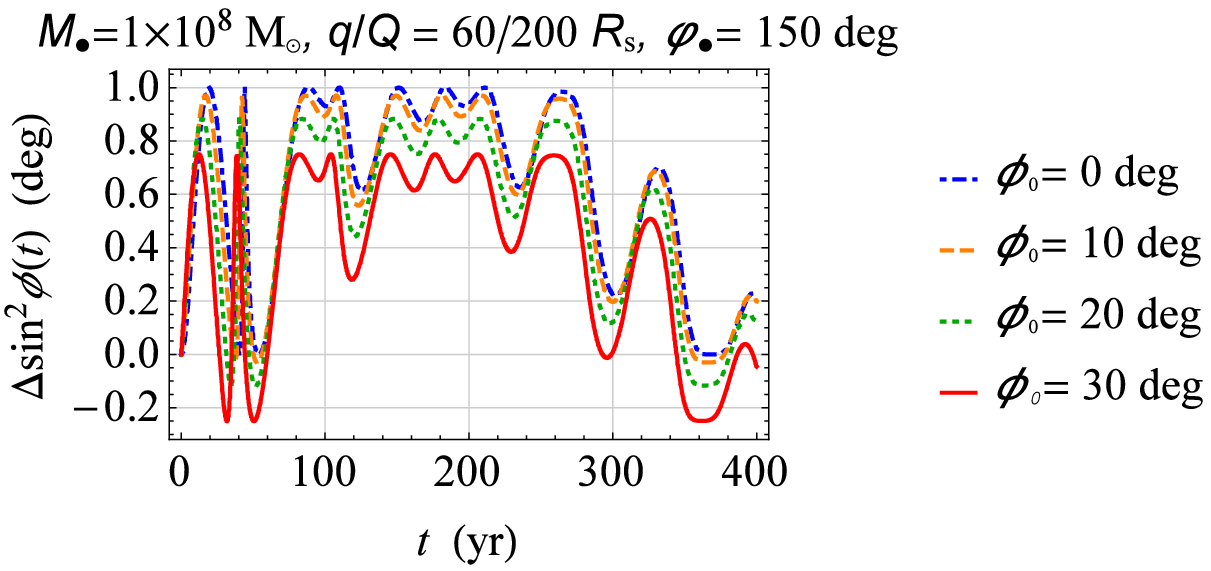} & \epsfxsize= 7.8 cm\epsfbox{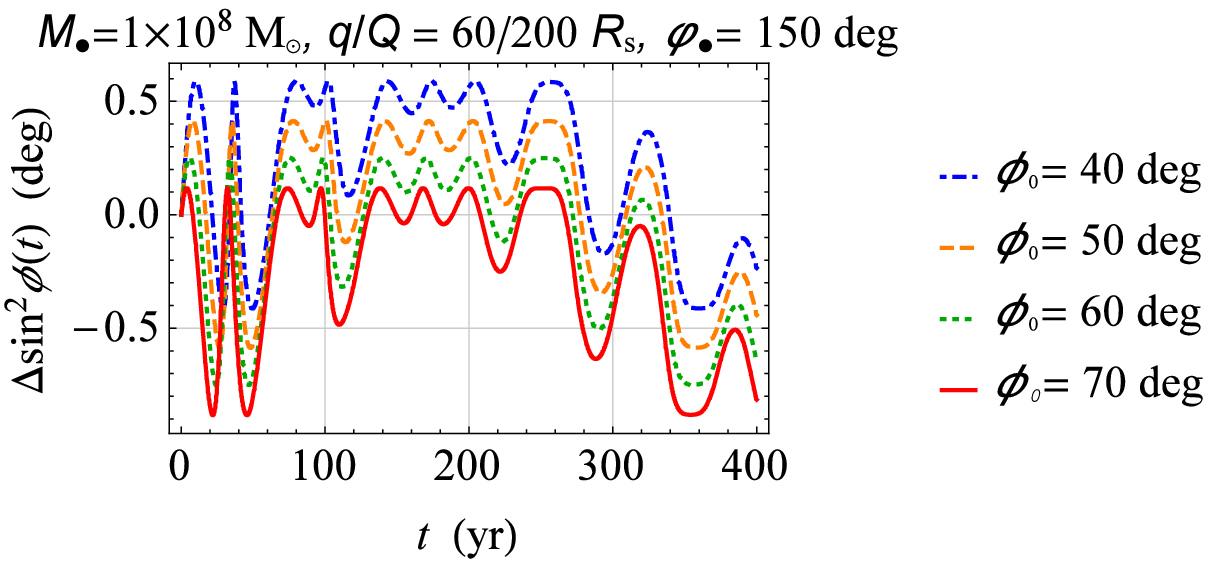}\\
\epsfxsize= 7.8 cm\epsfbox{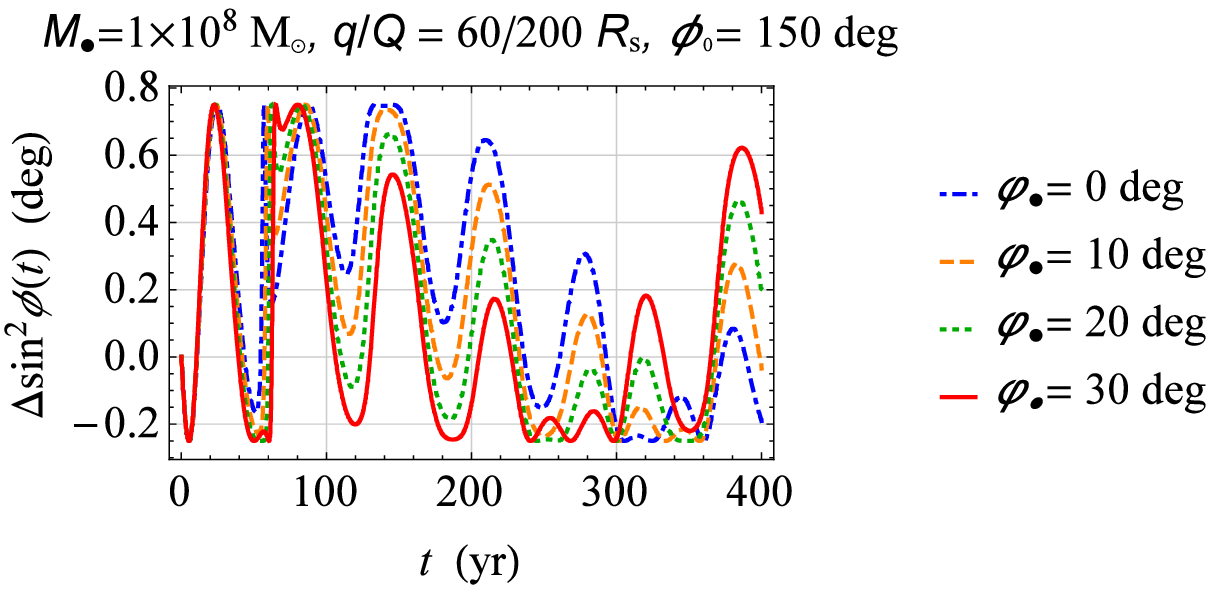} & \epsfxsize= 7.8 cm\epsfbox{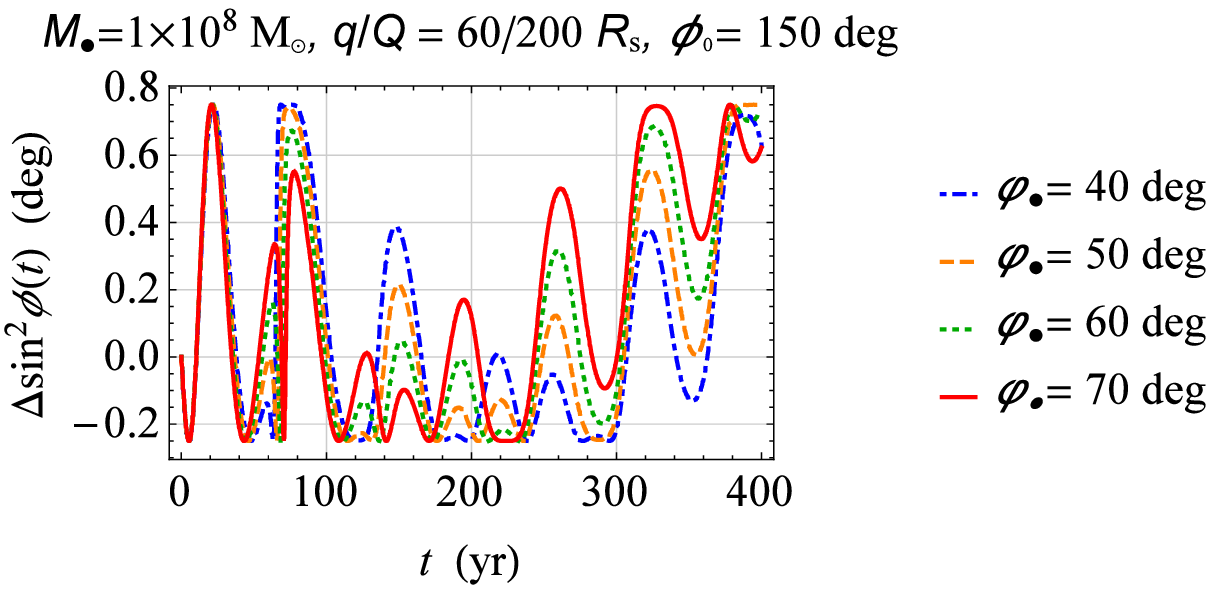}\\
\end{tabular}
}
}
\caption{
Numerically integrated time series of the shift $\Delta\sin^2\phi\ton{t}$, where $\phi(t)$ is the precession angle of the planet's spin axis, for different values of some parameters by assuming $M_\bullet = 1\times 10^8\,M_\odot,\,q = 60\,R_\mathrm{s},\,Q=200\,R_\mathrm{s}$, and generic initial conditions.  First and second rows: sensitivity to the obliquities $\eta_\bullet,\,\varepsilon_0$ of the spin axes of the SMBH and of the planet, respectively, by assuming  $\varphi_\bullet=150^\circ,\,\phi_0=150^\circ$ for their azimuthal angles. Third and fourth rows: sensitivity to the azimuthal angles $\varphi_\bullet,\,\phi_0$ of the spin axes of the SMBH and of the planet, respectively, by assuming $\eta_\bullet = 50^\circ,\,\varepsilon_0 = 23^\circ .4$ for their obliquities.}\label{figura5}
\end{figure}

\begin{figure}[H]
\centering
\centerline{
\vbox{
\begin{tabular}{cc}
\epsfxsize= 7.8 cm\epsfbox{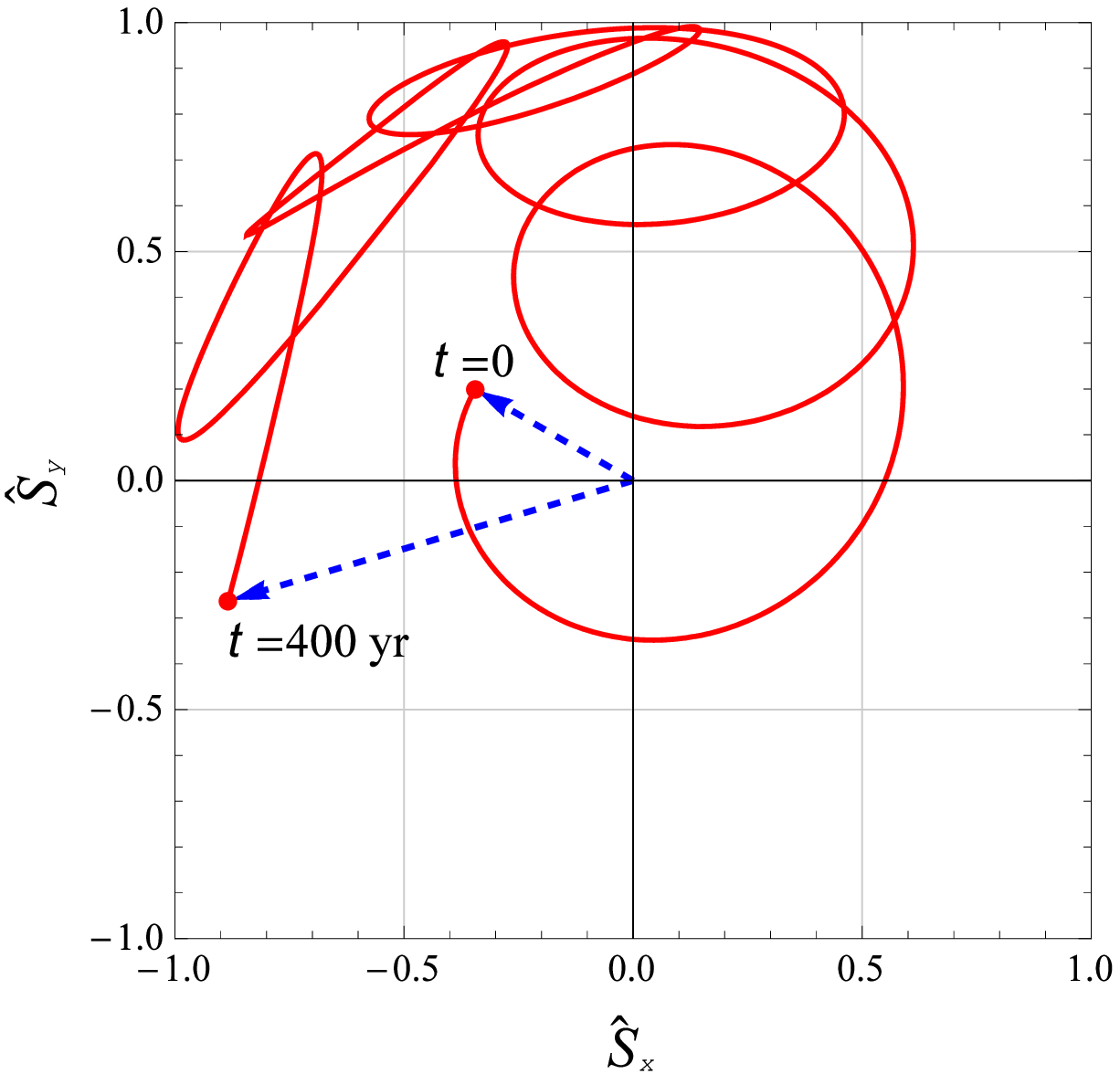} & \epsfxsize= 7.8 cm\epsfbox{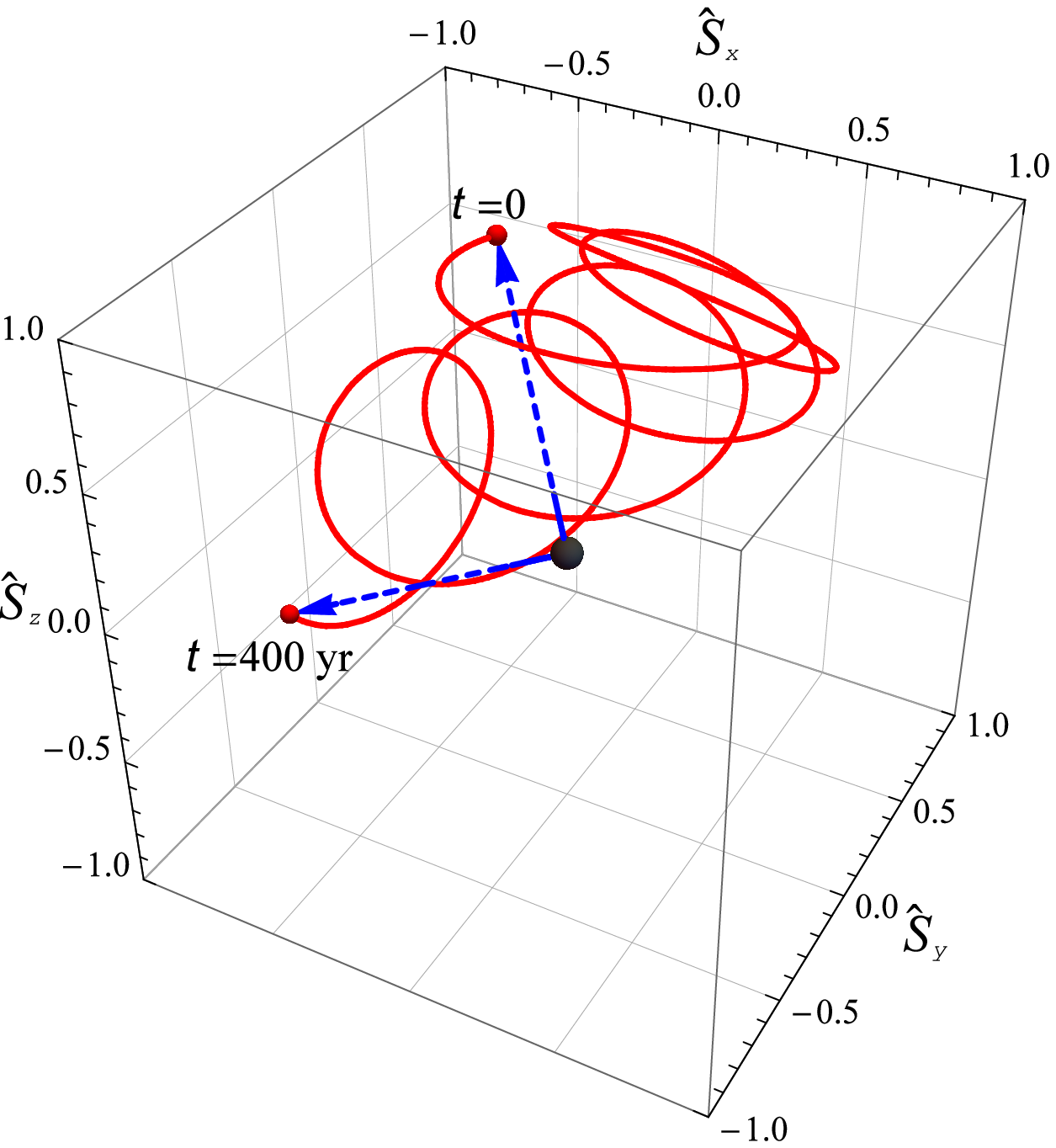}\\
\end{tabular}
}
}
\caption{
Left panel: numerically integrated hodograph, in red, of the planet's spin axis projection, represented by the dashed blue arrow, onto the $\grf{{\hat{S}}_x,\,{\hat{S}}_y}$ plane over $\Delta t=400\,\mathrm{yr}$. Right panel: numerically integrated hodograph, in red, of the planet's spin axis (dashed blue arrow) over the same time span. In both cases, a maximally rotating SMBH with $M_\bullet = 1\times 10^8\,M_\odot,\,\eta_\bullet=50^\circ,\,\varphi_\bullet=150^\circ$ is assumed. The initial conditions of the spin axis of the planet, which orbits from an aponigricon $Q=200\,R_\mathrm{s}$ to a perinigricon $q=60\,R_\mathrm{s}$, are $\varepsilon_0=23^\circ .4,\,\phi_0=150^\circ$. Cfr. with the dashed--dotted blue curve in the upper right panel of Figure\,\ref{figura4}.}\label{figura6}
\end{figure}

\lb{altri}
Now, we will explore a little more the parameter space by allowing for a much more eccentric and slower orbital motion of the fictitious planet characterized by a perinigricon distance $q=60\,R_\mathrm{s}$ and an aponigricon distance $Q=200\,R_\mathrm{s}$. Figure\,\ref{figura4} and Figure\,\ref{figura5} show the obliquity and precessional shifts of the planet's spin axis. It can be noticed that the huge eccentricity somewhat compensates the damping effect of the larger semimajor axis because the magnitude of the obliquity's signals is similar to that in Figure\,\ref{figura1} for $e=0,\,r=100\,R_\mathrm{s}$. The sensitivity to the parameters of both the spin axes turns out to be as for the circular orbit case.
Figure\,\ref{figura6} shows the numerically integrated hodograph of the planet's spin axis of one of the numerical integrations
of the upper right panels of Figure\,\ref{figura4} and Figure\,\ref{figura5}.

\begin{figure}[H]
\centering
\centerline{
\vbox{
\begin{tabular}{cc}
\epsfxsize= 7.8 cm\epsfbox{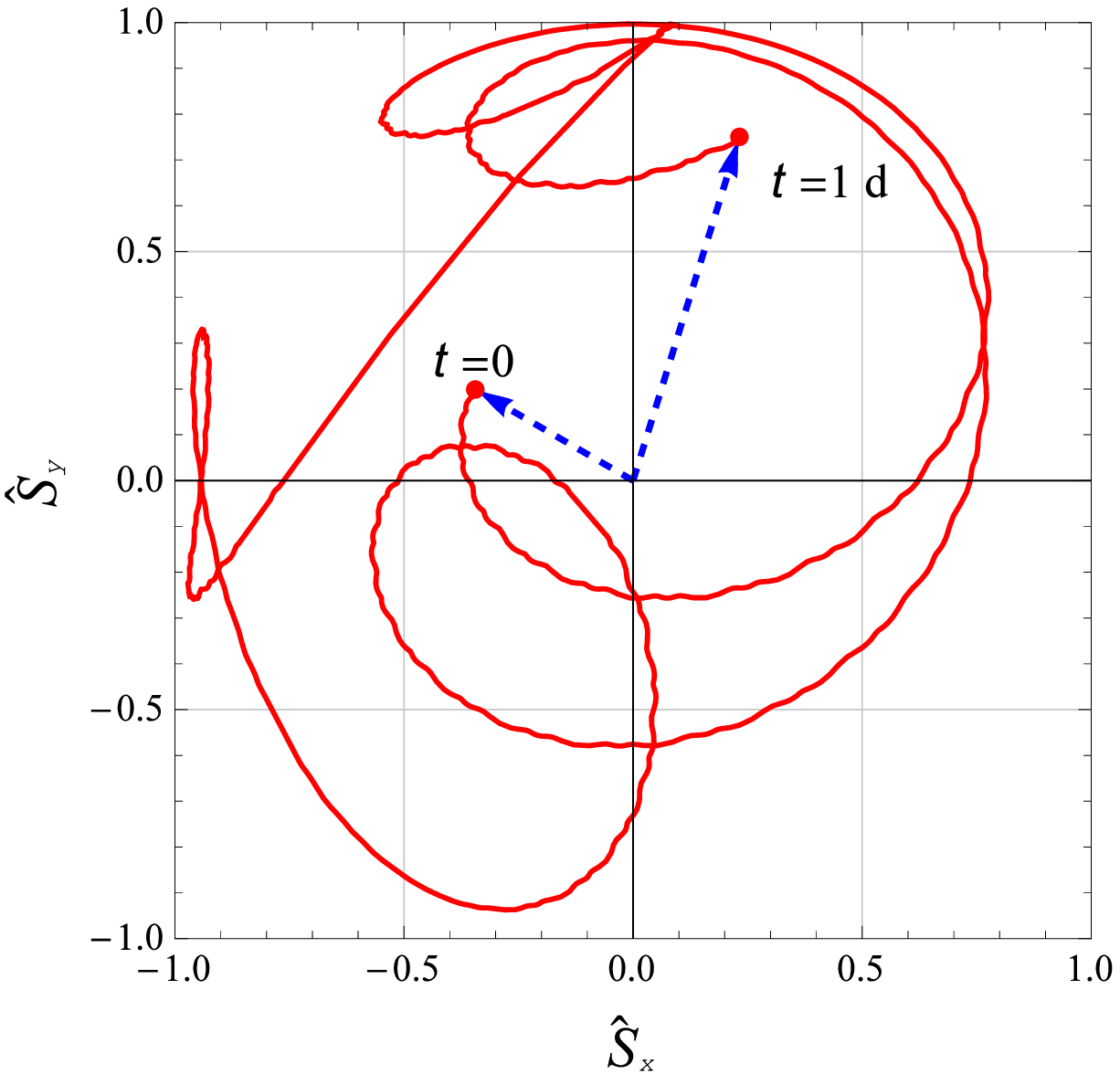} & \epsfxsize= 7.8 cm\epsfbox{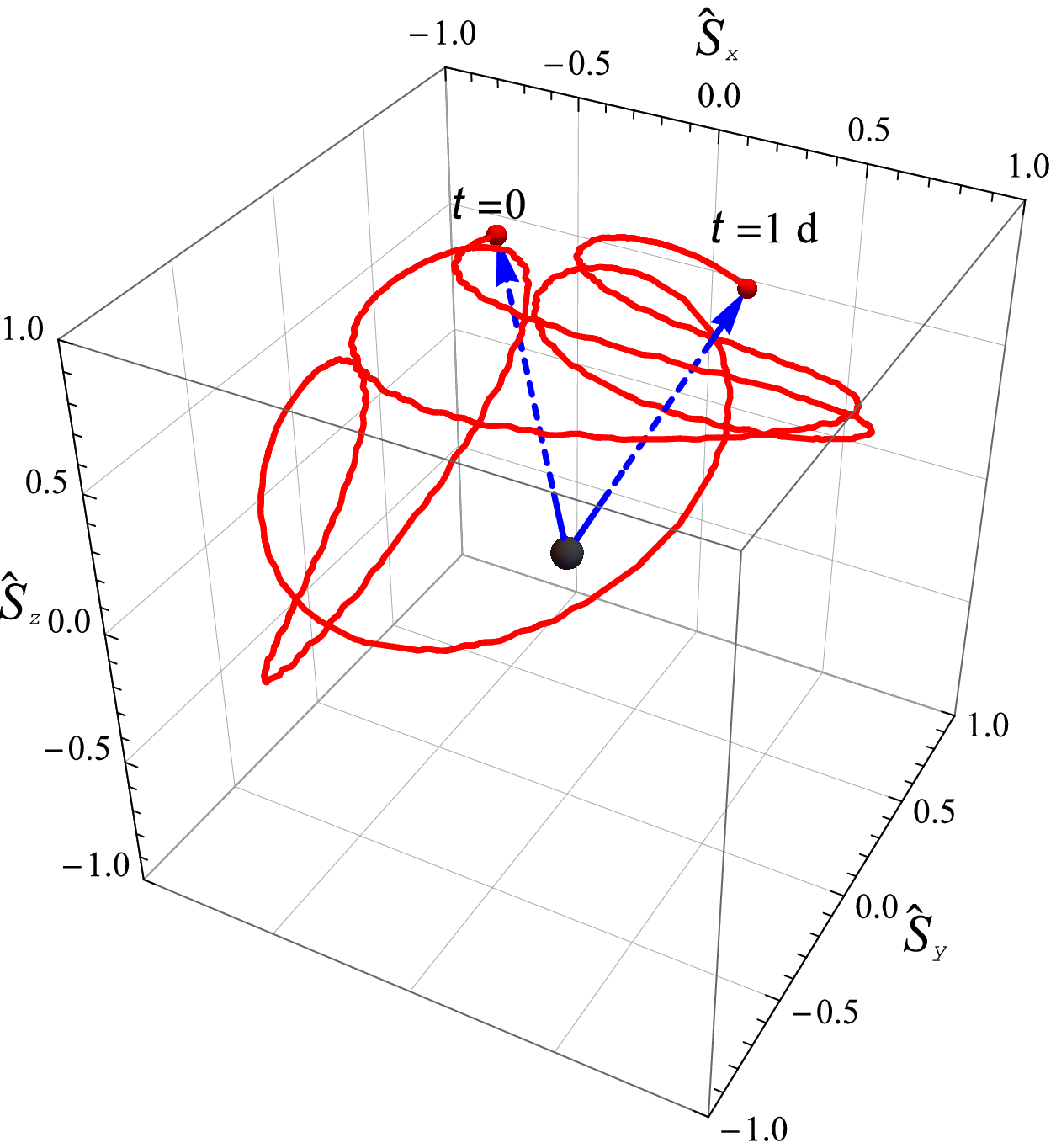}\\
\end{tabular}
}
}
\caption{
Left panel: numerically integrated hodograph, in red, of the planet's spin axis projection, represented by the dashed blue arrow, onto the $\grf{{\hat{S}}_x,\,{\hat{S}}_y}$ plane over $\Delta t=1\,\mathrm{day}$. Right panel: numerically integrated hodograph, in red, of the planet's spin axis (dashed blue arrow) over the same time span. In both cases, a maximally rotating SMBH with $M_\bullet = 1\times 1\times 10^5\,M_\odot,\,\eta_\bullet=50^\circ,\,\varphi_\bullet=150^\circ$ is assumed. The initial conditions of the spin axis of the planet, which orbits from an aponigricon $Q=15\,R_\mathrm{s}$ to a perinigricon $q=10\,R_\mathrm{s}\,(e=0.2)$, are $\varepsilon_0=23^\circ .4,\,\phi_0=150^\circ$. }\label{figura7}
\end{figure}

Here, in order to further dig into the parameters space, we will change also the mass of the SMBH by adopting $M_bullet=1\times 10^5\,M_\odot$ for it.
We will look at a rather eccentric trajectory $(e=0.2)$ delimited by $q=10\,R_\mathrm{s},\,Q=15\,R_\mathrm{s}$. To give an idea of the essential features of this scenario, we will perform just one run by using the same initial conditions for both the spin axes used in some of the previous cases. Figure\,\ref{figura7} displays the numerically integrated trajectory of the planet's spin axis over a time span $\Delta t = 1\,\mathrm{day}$.
\section{Summary and Conclusions}\lb{fine}
Recently, the possibility that a huge number of rogue telluric planets, potentially able to sustain Earth-like life under certain circumstances, may form in the neighborhood of SMBHs  has gained growing attention in the literature. Several physical phenomena of different nature have been considered so far in connection to their role in favoring or disadvantaging the emergence of habitable ecosystems in such worlds that would receive the required energy input mainly from a possible accretion disk surrounding the SMBH.
Among the various scenarios appeared in the literature, an interesting one consists of a putative planet orbiting a SMBH at such a distance, $r$, that its lensed disk would appear to an observer on the planet orbiting just outside the SMBH's equatorial plane as large as the Sun as seen from the Earth; for $M_\bullet = 1\times 10^8\,M_\odot$, it would be $r=100\,R_\mathrm{s}$. For a certain mass accretion rate, it would even be possible to set the disk's temperature equal to the solar one.

Here, we demonstrated that among the many pros and cons that must be weighed in a plausible assessment of the habitability of such a world, there is also the effect that general relativity exerts on the tilt $\varepsilon$ of its spin axis to the \virg{ecliptical} plane. Indeed, by numerically integrating the planet's pN equations of motion along with the pN evolution equations of its spin axis, it turned out that its obliquity $\varepsilon$ may experience remarkable changes $\Delta\varepsilon\ton{t}$ over a comparatively short time, mainly depending on the obliquity $\eta_\bullet$ of the spin axis of the SMBH, assumed maximally rotating, and, although to a lesser extent, on the initial value $\varepsilon_0$ of the planet's obliquity itself. Indeed, $\Delta\varepsilon\ton{t}$ undergoes oscillating variations over a time span, say, $\Delta t= 400\,\mathrm{yr}$ whose size may be as large as tens or even hundreds of degrees. The largest effects occur for $\eta_\bullet$ approaching $90^\circ$, but also if the SMBH's spin axis is nearly perpendicular to the orbital plane $(\eta_\bullet \simeq 10^\circ-20^\circ)$ the amplitude of the change of the planet's obliquity can reach the $\simeq 20^\circ-40^\circ$ level.
%

The impact of a large orbital eccentricity and different distance from the SMBH were investigated as well. We found that, for, say, a semimajor axis of $130\,R_\mathrm{s}$ and $e = 0.538$, some sort of compensation occurs since the size of the signals of $\Delta\varepsilon\ton{t}$ does not change too much.

I am grateful to the anonymous referee for her/his helpful remarks.
\bibliography{MR_biblio,MS_binary_pulsar_bib,Gclockbib,semimabib,PXbib}{}


\end{document}